\documentclass[lettersize,journal]{IEEEtran}
\usepackage{amsmath,amsfonts}
\usepackage{algorithmic}
\usepackage{algorithm}
\usepackage{array}
\usepackage[caption=false,font=normalsize,labelfont=sf,textfont=sf]{subfig}
\usepackage{textcomp}
\usepackage{stfloats}
\usepackage{url}
\usepackage{verbatim}
\usepackage{graphicx}
\usepackage{cite}
\usepackage{makecell}

\usepackage{multirow, color, rotating, subcaption}
\usepackage{hyperref}
\begin{document}

\title{A Scalable Hierarchical Intrusion Detection System for\\ Internet of Vehicles}

\author{Md Ashraf Uddin, Nam H. Chu, Reza Rafeh, and Mutaz Barika
% \author{Nam Hoai Chu,~\IEEEmembership{Staff,~IEEE,}  % Ashraf Uddin- ashraf.uddin@cihe.edu.au   Nam Hoai Chu - namhoai.chu@cihe.edu.au
% \author{Reza Rafeh,~\IEEEmembership{Staff,~IEEE,}     reza.rafeh@cihe.edu.au
% \author{Mutaz Barika,~\IEEEmembership{Staff,~IEEE,}   mutaz.barika@cihe.edu.au

        % <-this % stops a space
\thanks{The authors are with the School of Information Technology, Crown Institute of Higher Education, Australia. (emails: \{\textit{ashraf.uddin, namhoai.chu, reza.rafeh, mutaz.barika\}@cihe.edu.au})}% <-this % stops a space
\thanks{Manuscript received XX XX, 2024; revised XX YY, 2025.}}

\markboth{Journal of \LaTeX\ Class Files,~Vol.~XX, No.~XX, XX~2025}%
{Shell \MakeLowercase{\textit{et al.}}: A Sample Article Using IEEEtran.cls for IEEE Journals}


\maketitle

\begin{abstract}
Due to its nature of dynamic, mobility, and wireless data transfer, the Internet of Vehicles (IoV) is prone to various cyber threats, ranging from spoofing and Distributed Denial of Services (DDoS) attacks to malware. 
    To safeguard the IoV ecosystem from intrusions, malicious activities, policy violations, intrusion detection systems (IDS) play a critical role by continuously monitoring and analyzing network traffic to identify and mitigate potential threats in real-time. 
    However, most existing research has focused on developing centralized, machine learning-based IDS systems for IoV without accounting for its inherently distributed nature. 
    Due to intensive computing requirements, these centralized systems often rely on the cloud to detect cyber threats, increasing delay of system response. 
    On the other hand, edge nodes typically lack the necessary resources to train and deploy complex machine learning algorithms.
    To address this issue, this paper proposes an effective hierarchical classification framework tailored for IoV networks. 
    Hierarchical classification allows classifiers to be trained and tested at different levels, enabling edge nodes to detect specific types of attacks independently. 
    With this approach, edge nodes can conduct targeted attack detection while leveraging cloud nodes for comprehensive threat analysis and support. 
    Given the resource constraints of edge nodes, we have employed the Boruta feature selection method to reduce data dimensionality, optimizing processing efficiency. 
    % In addition, we implemented and evaluated a distributed random forest model using the H2O framework, where each tree in the forest can be processed by an edge node, allowing collaborative attack detection across the network. 
    % Addressing the security and privacy challenges inherent in distributed IoV networks, we also trained and tested a federated deep neural network (DNN) model using the Flower framework. 
    To evaluate our proposed framework, we utilize the latest IoV security dataset CIC-IoV2024, achieving promising results that demonstrate the feasibility and effectiveness of our models in securing IoV networks. 
\end{abstract}

\begin{IEEEkeywords}
Internet of Vehicles; Network traffic; Intrusion Detection System; Machine Learning; Hierarchical Classification; Flat Classification.
\end{IEEEkeywords}

\section{Introduction}
\label{section:introduction}
\IEEEPARstart{T}{he} Internet of Vehicles (IoV) is rapidly growing thanks to recent breakthrough in technologies.
    Specifically, 5G offers a wide range of use cases to facilitate IoV system, e.g., ultra-reliable low latency communications (uRLLC), enhanced mobile broadband (eMBB), and massive machine type communications (mMTC).
    On the other hand, recent advances in sensors, cameras, and Electronic Control Units (ECU) enable the new generation of smart vehicles, improving comfort and safety for drivers and passengers. 
    Technically, IoV is a part of Intelligent Transportation System (ITS) and combines two concepts: Vehicular Ad Hoc Networks (VANET) and Internet of Things (IoT) \cite{taslimasa2023security}.
    By leveraging IoT, vehicles now can connect to the Internet and receive real-time data about traffic, hazardous, navigation, etc. According to \cite{goasduff202023gartner}, by 2030, the automotive industry will use 53\% of the 5G IoT endpoints. 

\begin{figure}[!t]
   \centering
    \includegraphics[width=0.9\linewidth]{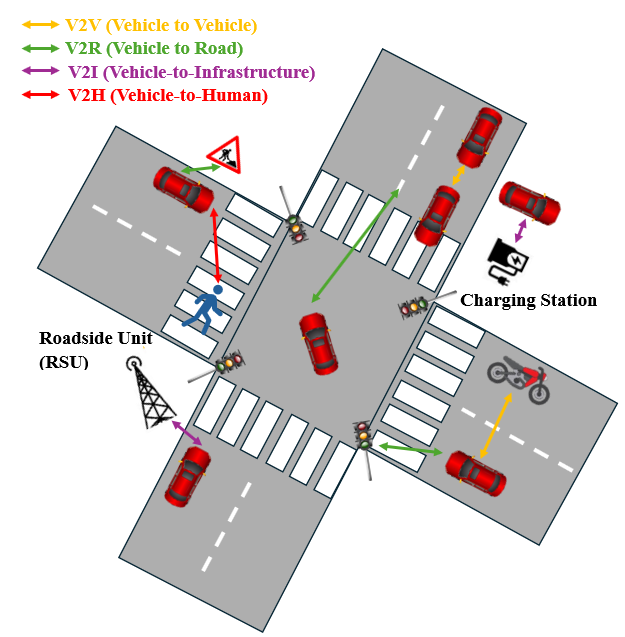}
    \caption{\textcolor{black}{IoV communications}}
    \label{fig:iov_communication}
    \vspace{-20 pt}
\end{figure} 

Generally, IoV communications are divided into two categories: Intra-Vehicle and Inter-Vehicle, as shown in~ Fig. \ref{fig:iov_communication}. 
    In particular, Intra-Vehicle is responsible for all communications between components (e.g., sensors, cameras, and ECUs) inside a vehicle. 
    This exchanged data is usually used for autonomous driving, entertainment, and advisory information. 
    On the other hand, Inter-Vehicle facilitates all communications between cars and its surrounding entities, such as other cars, pedestrians, roadside units (RSU), and grids. 
    This exchanged data via IoV communications is usually used for autonomous driving, and traffic management (e.g., road safety, and navigation and advisory information).    
    % Thus, securing IoV communications is crucial for the operation of an IoV system.
    {\color{black}Typically, IoV data is often processed across three layers \cite{talebkhah2024task}: (i) \textit{Vehicle layer} that processes local data to identify some hazardous situations and make suitable decisions to control the vehicle or inform the driver, (ii) \textit{Edge layer} consisting devices like RSUs that communicates with vehicles to update and promptly response to vehicles requests (e.g., navigation and traffic management), and (iii) \textit{Cloud layer} that has enough resource capacity to handle intensive tasks, such as analysing, aggregating and storing traffic data.

% Reviwewer 1 - Q1
Given the above, the involvement of numerous heterogeneous entities (from sensors, vehicles, RSUs/Edge, to cloud), various wireless technologies (from short range, e.g., Bluetooth, to long range, e.g., 5G), and node mobility makes IoV vulnerable to many types of attacks.     
    For example, intruders can intercept and tamper with data to gain control of vehicles, mislead the control systems, and perform other types of malicious activities. 
    Since any malicious activities in IoV can be life-threatening, ensuring its security is of utmost importance~\cite{taslimasa2023security}.    
    In 2015, two cybersecurity researchers demonstrated a vulnerability in the Jeep Cherokee using its Uconnect infotainment system, which allowed them to take control of the engine, brakes, acceleration, and other functions. They exploited an open port in the Uconnect system and used a zero-day exploit to send commands through the Jeep’s entertainment system to its dashboard functions, effectively gaining control of the entire car \cite{miller2015remote}.
    }

However, it is challenging or even infeasible to protect edge nodes with traditional security measures like firewalls, encryption, and anti-malware tools, as these nodes often lack computing capabilities.
% continuously face new and unknown threats, i.e., zero-day attacks \cite{mirzaee2021two}. 
   In addition, although conventional IDS that can identify diverse cyberattacks are essential for IoV, their effectiveness can vary based on data sources, operational models, and response mechanisms. 
    For example, a network-based IDS monitors network traffic from connected devices/vehicles, while a host-based IDS focuses on detecting threats originating from individual user devices \cite{mohy2022effective}. 
    Meanwhile, Signature-Based IDS (SIDS) detect threats by matching network traffic patterns against known attack signatures stored within the system. 
    In contrast to the above approach, Anomaly-Based IDS (AIDS) leverage probabilistic models to detect previously unknown or zero-day attacks. 
    By training on large, relevant IDS datasets, AIDS learn to analyse network traffic patterns and system behaviours, enabling it to identify deviations indicative of potential cyberattacks \cite{mohy2022effective}. 
    In the dynamic environment of IoV, where real-time threat detection is critical, the adaptability of Anomaly-Based IDS provide a significant advantage by offering proactive detection capabilities against novel threats and minimising the reliance on predefined signatures.
    
As such, in the IoV-IDS literature, Anomaly-Based IDS has been adopted in various classification approaches to differentiate between normal traffic and multiple types of cyberattacks \cite{anbalagan2023iids, aloraini2024adversarial, yang2021mth, neto2024ciciov2024}.  
    Although these approaches use different classification algorithms, the majority rely on a single, flat multi-class classifier to simultaneously distinguish normal traffic from various attack types. 
\textcolor{black}{However, this flat classification approach is not ideal for the IoV network for the following reasons.
    \begin{itemize}
        \item IoV is a dynamic, distributed, and resource-constrained environment where edge nodes (e.g., roadside units) lack the computational resources necessary to train or retrain a flat or centralized classifier using large datasets. 
        \item IoV network data is often highly imbalanced~\cite{kostage2024enhancing}, with the majority of traffic being normal and only a small fraction representing attack types. 
        A flat classifier trained on such unbalanced data tends to be biased toward the normal class, resulting in a high false-negative rate where attacks are misclassified as normal traffic. 
        In network security, this issue can have catastrophic consequences. 
    % {\color{red}Therefore, minimising false negatives is a key objective, while still ensuring legitimate requests have unobstructed access to the network.}
        \item Certain IoV attack types may exhibit similarities, making them difficult to be distinguished with a single, global classifier.
        \item The highly distributed and dynamic characteristics of IoV environments make flat classification models inefficient.
        \item Scalability is a significant challenge in flat IDS architectures. 
        As the number of vehicles increases, a centralized IDS must handle an overwhelming volume of requests, which can lead to bottlenecks and introduce a single point of failure. 
        This dependency on a centralized system also creates dependencies on network resources, which are not uniformly available across different geographical areas, further complicating communication with the centralized IDS. 
        These limitations make it difficult to be effectively deployed in IoV networks over large geographic regions.
    \end{itemize}
Thus, it necessitates an innovative approach that can address the above issues effectively.   
In this context, hierarchical classification models have been emerging as the potential solutions for securing IoV networks. 
    The main reason is that hierarchical models allow for targeted, layered classification of attack types.
    By doing so, it can reduce false positives and negatives, thereby improving overall attack detection accuracy. 
    % Thus, it can overcome the limitations of traditional flat classifiers and support more effective and efficient cyber threat detection in IoV environments. 
    Additionally, this approach enables edge nodes to train/retrain machine learning models in a decentralized manner, using only the resources available locally. 
    Specifically, individual vehicles can detect and respond to cyberattacks using decentralized IDS mechanisms, eliminating the need to rely solely on a centralized IDS. 
    This distributed architecture not only reduces the risk of bottlenecks and failure but also ensures that IDS functionalities can operate efficiently even in areas with limited network connectivity. 
    Given the above, the hierarchical model enhances the resilience and scalability of the IoV network, enabling it to support larger, geographically dispersed deployments.     }
    % By doing so, a hierarchical classification approach offers a promising solution to these issues of scalability and availability.    
    Therefore, in this study, we propose a scalable three-level hierarchical model for classifying various attacks within an Internet of Vehicles (IoV) network. 

{\color{black}The contributions of our work are summarised as follows:
   \begin{itemize}
        \item Design a scalable hierarchical IDS framework considering resource constrains of IoV. Specifically, the proposed framework consists of three level classifications, including (i) Level 1: a binary classifier differentiates between normal and attack instances, (ii) Level 2: a multi-class classifier categories an attack into broad categories, and (iii) Level 3: multiple multi-class classifiers identify specific subcategories of attacks, providing detailed classifications. 
        \item Propose an effective feature selection method based on Boruta to minimize training and testing complexity.
        \item Comprehensively evaluate the performance of the proposed approach on the most recent IoV security dataset CIC-IoV2024.
        \item Provide insights of the CIC-IoV2024 dataset, highlighting the most important features in this dataset.
    \end{itemize}}

The structure of this paper is as follows. Section \ref{section:Literature_Review} presents a review of the related literature. Section \ref{Proposed} details the hierarchical classification architecture and materials used in this study. In Section \ref{section:RESULTS AND DISCUSSION}, we present the results of our experiments conducted to compare them with their flat counterparts. Finally, Section \ref{Conclusion} summarizes the paper and outlines potential future research directions.

%======================================
\section{Related works} 
\label{section:Literature_Review}

This section will analyze existing works closely related to our research. Specifically, it first provides an overview of recent IDS solutions based on hierarchical architecture. Then, it presents the state-of-the-art IDS for IoV networks. Finally, the main contributions of this work are highlighted.

\subsection{Hierarchical Classification in IDS}

IDS plays a vital role in safeguarding network infrastructures, and hierarchical classification approaches have emerged as promising techniques to improve detection accuracy. 
    A variety of hierarchical models have been proposed to tackle the multi-classification challenges posed by diverse attack types. 
    Some studies focused on two-level hierarchical models to improve attack classification~\cite{alin2020towards, sarnovsky2020hierarchical, ahmim2019novel, Uddin2024, UDDIN2025108006}. 
    Alin et al. (2020)~\cite{alin2020towards} proposed a two-level IDS where the first level employed a binary classifier to distinguish between benign and malicious traffic, while the second level used a multi-classifier to identify specific attack types. 
    In~\cite{sarnovsky2020hierarchical}, Sarnovsky et al. (2020) adopted a similar two-level hierarchical approach but using an ensemble classifier in the second level. 
    While the approaches in~\cite{alin2020towards, sarnovsky2020hierarchical} train classifiers independently, Ahmim et al. (2019) developed a two-level model that utilized outcomes from first-level classifiers to train second-level classifiers on datasets with original labels~\cite{ahmim2019novel}.     
    Different from the above works~\cite{alin2020towards, sarnovsky2020hierarchical, ahmim2019novel}, Uddin et al. focus on addressing the zero-day attacks~\cite{Uddin2024, UDDIN2025108006}. 
    The study~\cite{Uddin2024} proposes a dual-tier adaptive IDS, where the first tier leverages one-class classification (OCC) to distinguish normal activities. 
    In the second tier, it employs multi-class classification to identify whether the detected threats are known or unknown. 
    In~\cite{UDDIN2025108006}, the study aim to address the challenge of detecting zero-day attacks without requiring real attack samples by using synthetic attack data.
    Notably, the common shortcoming of the above works is that they only leverage two tiers, possibly resulting in reduced accuracy when classifying more complex attack categories, such as those involving three levels of attack types.
    
Aim to further enhance the attack detection accuracy, several works consider three or more levels of classifiers, e.g.,~\cite{sarikaya2020class, mohd2021intrusion, eldahshan2022meta, alzaqebah2023hierarchical, verkerken2023novel}. 
    In~\cite{sarikaya2020class}, Sarikaya et al. (2020) proposed a three-stage hierarchical classification model to detect network intrusions. 
    The first stage involved a binary Random Forest (RF) classifier to distinguish between benign and malicious traffic. In the second stage, attacks were grouped into two clusters (i.e., Group 1: DoS and Exploit attacks, and Group 2: other attacks) {based on the confusion matrix analysis of a flat classifier}. 
    A binary RF classifier was employed for each group. 
    The third stage utilised a multi-class RF classifier to identify specific attack types within Group 2. 
    
    Unlike the proposed solution in~\cite{sarikaya2020class}, Mohd et al. (2021) introduced a hierarchical classification model based on a One-vs-Rest (OvR) strategy~\cite{mohd2021intrusion}. 
    Their approach involved training five classifiers—SVM, Probabilistic Neural Network (PNN), Decision Tree (DT), Neuro-fuzzy Classifier (NFC), and Smooth Support Vector Machine (SSVM)—{with each classifier trained to detect four distinct attack types at four hierarchical levels, as well as normal traffic}. 
    Then, the best-performing classifiers were selected at each level to construct the final model.  
    The OvR approach is advanced by integrating optimisation techniques by Eldahshan et al. (2022)~\cite{eldahshan2022meta}. 
    Their methodology involved a three-stage pipeline comprising Binary Grey Wolf Optimizer (GWO) feature selection, an Extreme Learning Machine (ELM), and metaheuristic-based hyperparameter tuning using the Archimedes Optimization Algorithm (AOA) and Honey Badger Algorithm (HBA). 
    Following this direction, Alzaqebah et al. (2023) presenting a hierarchical IDS model incorporating bio-inspired feature selection~\cite{alzaqebah2023hierarchical}. 
    Their model used an ELM to classify attacks at different levels, with an enhanced Harris Hawks optimizer (IHHO) optimizing the feature set, weights, and biases.
    Recently, Verkerken et al.~\cite{verkerken2023novel} proposed a hierarchical intrusion detection that includes three stages and leverages Deep Learning together with traditional machine learning mechanism. 
    In the first stage, an anomaly detector is used to identify suspicious records. 
    They used an autoencoder (AE) and one class support vector machine (OC-SVM) for this purpose. 
    In the second stage, a RF and a neural network (NN) are used to classify suspicious records from the first stage into multi-class known attacks. 
    The third stage whose purpose is to reduce the false positive rate uses the anomaly score outputted from the first stage to identify remaining attack records. %The best accuracy of 96.08\% was achieved using AE for the first stage and RF for the second stage and the threshold of 0.95 for the third stage.

All of the above studies (i.e.,~\cite{alin2020towards, sarnovsky2020hierarchical, ahmim2019novel, Uddin2024, UDDIN2025108006, sarikaya2020class, mohd2021intrusion, eldahshan2022meta, alzaqebah2023hierarchical, verkerken2023novel}) regarding hierarchical classification methods for IDS only consider a typical network and IoT attacks, leaving the real-world IoV network attacks unexplored. 
    Additionally, their proposed frameworks may not work effectively in IoV system due to several problems. For example, they may require intensive computing resources (e.g.,~\cite{verkerken2023novel, eldahshan2022meta, alzaqebah2023hierarchical}), which are challenging or even impossible to be provided by IoV devices. 
    Some studies use outdated dataset (e.g.,~\cite{alin2020towards, sarnovsky2020hierarchical, mohd2021intrusion}) or rely heavily on manual grouping~\cite{sarikaya2020class}, limiting the generalizability of their findings and the applicability in practice.  
    The next subsection will analyze the state-of-the-art studies regarding IDS in IoV networks.

\subsection{IDS Datasets for IoV Networks}

Although numerous IoT security datasets are available, datasets specifically focused on IoV security remain limited.
    As such, many works regarding IoV security in the literature utilize general IoT security datasets or even typical network security datasets.
    For example, the studies \cite{kumar2021p2sf, gad2021intrusion, kumar2021privacy, abdel2021federated} use IoT BotNet \cite{ullah2020technique} and ToNIoT~\cite{moustafa2021new} dataset, which are collected from typical IoT attacks. 
    On the other hand, other non-IoV security datasets (e.g., NSL-KDD~\cite{tavallaee2009detailed}, UNSW-NB15~\cite{moustafa2015unsw}, and AWID~\cite{chatzoglou2021empirical}, CIC-DDoS-2019~\cite{sharafaldin2019developing}) have been utilized in studies like in~\cite{zhang2018distributed, li2021transfer, jin2021intrusion, korium2024intrusion}. 
    The evaluations on non IoV security datasets make these studies (i.e.,~\cite{kumar2021p2sf, gad2021intrusion, kumar2021privacy, abdel2021federated,zhang2018distributed, li2021transfer, jin2021intrusion, korium2024intrusion}) less applicable for IoV environments.

Recently, several IoV security datasets collected from real vehicles have been introduced, including GIDS~\cite{seo2018gids}, OTIDS~\cite{lee2017otids}, Road~\cite{verma2020road}, and CIC-IoV2024~\cite{neto2024ciciov2024}. 
    Among these datasets, CIC-IoV2024’s support for diverse data representations and comprehensive attack coverage makes it an ideal choice for machine learning-based IDS for IoV. 
    Therefore, this dataset has been leveraged in the most recent IoV security studies\cite{aswal2024advancing, gul2024improving, merzouk2024adversarial, yagiz2024transforming, mahdi2024advanced}. 
{\color{black}Specifically, Aswal et al. propose a typical Deep Neural Network (DNN)-based detector, which consists of three fully connected layers, and evaluate it on the CICIoV2024 dataset~\cite{aswal2024advancing}.
    Similarly, Gul et al. propose a Machine Learning (ML)-based IDS for the IoV system~\cite{gul2024improving}.
    In their framework, the Generic Algorithm (GA) is used to tune the parameters of ML algorithms, including Logistic Regression (LR), RF, AdaBoost, and DNN.   
    Differently, other works leverage more advanced ML techniques, e.g., Deep Reinforcement Learning (DRL), Variational Autoencoder (VAE), and Long Short-term Memory (LSTM) ~\cite{merzouk2024adversarial, yagiz2024transforming, mahdi2024advanced}.
    In particular, Merzouk et al.~\cite{merzouk2024adversarial} leverage DRL to identify the attacks.
    However, it only classifies an event into the attack or benign groups, making it less applicable for a comprehensive attack detection.
    On the other hand, Yagiz et al.~\cite{yagiz2024transforming} proposed an IDS based on the VAE architecture and Knowledge Distillation (KD) technique. 
    Particularly, it uses an encoder to map the input data to a latent space and then use a decoder to reconstruct the input data from the latent space. 
    The reconstruction error indicates which input data are not normal. 
    Their proposed model also used transfer learning through knowledge distillation where the knowledge from a large model (teacher) is transferred to a smaller model (student). 
    Whereas, in~\cite{mahdi2024advanced}, Mahdi et al. proposed a hybrid approach combining LSTM architecture with a traditional ML, i.e, Naive Bayes classifier.
    However, these approaches~\cite{yagiz2024transforming, mahdi2024advanced} require intensive computing resources and a huge amount of memory, making it less applicable to IoV systems.
    As such, they do not leverage the entire dataset.
    To achieve a reasonable decision delay, the work~\cite{yagiz2024transforming} can only use a small portion of the dataset while~\cite{mahdi2024advanced} only use the spoofing attack data, possibly leading to a low data usage.
Table \ref{tab:litsummary} summarizes the literature findings and highlights the shortcomings of existing works in addressing IoV security needs.}

\begin{table*}[]
    \caption{\color{black}Summary of literature findings}
    \label{tab:litsummary}
    \centering
\begin{tabular}{|l|m{4cm}|m{4.5cm}|m{4cm}|l|}
\hline
\textbf{Literature} & \textbf{Attack Type(s)} & \textbf{Contribution} & \textbf{Dataset Used} \\ \hline
Zhang et al. (2018) [31] & Intrusion Attacks  & Distributed machine learning & NSL-KDD \\ \hline
Li et al. (2021) [32] & Attacks Against 802.11 & Transfer Learning & AWID \\ \hline
Jin et al. (2021) [33] & Analysis, Backdoor, DoS, Eploits, Fuzzers, Generic, Reconnaissance, Shellcode, Worms & Oversampling, outlier detection and metric learning & UNSW-NB15 \\ \hline
Korium et al. (2024) [34] & Dos, DDoS, Sniffing, Botnets, Infiltration, Web & RF, XGBoost, CatBoost, LightGBM & CIC-IDS-2017, CSE-CIC-IDS-2018, and CIC-DDoS-2019 \\ \hline
Seo et al. (2018) [35] & DoS, Fuzzy, RPM, GEAR & Generative Adversarial Nets & Car-Hacking \\ \hline
Lee et al. (2017) [36] & DoS and Fuzzy & Remote frame handling & OTIDS \\ \hline
Verma et al. (2020) [37] & Fabrication(DoS, Fuzzy, Targeted ID), suspension, Masquerade & A new dataset for CAN IDS& ROAD Dataset \\ \hline
Aswal et al. (2024) [38] & DoS and Spoofing& MLP-based deep learning &CIC-IoV2024 \\ \hline
Gül et al. (2024) [39] & DoS and Spoofing & RF and GA & CICIoV2024 \\ \hline
Merzouk et al. (2024) [40] & DoS and Spoofing & DRL & NSL-KDD, UNSW-NB15, and CIC-IoV2024 \\ \hline
Yagiz et al. (2024) [41] & DoS, Spoofing, and Fuzzing & SHAP+VAE+XAI, Transfer Learning & HCRL and  CIC-IoV2024 \\ \hline
Mahdi et al. (2024) [42] & DDoS and Spoofing & LSTM and Naive Bayes & CIC-DDoS2019 \\ \hline
Our Proposed Approach & DoS and Spoofing & Hierarchical classification  & CIC-IoV2024 \\ \hline
\end{tabular}
\end{table*}
% \vspace{-2pt}

Given the above, existing studies, regarding IDS for IoV~\cite{aswal2024advancing, gul2024improving, merzouk2024adversarial, yagiz2024transforming, mahdi2024advanced}, often consider a flat classifier, which is not effective in IoV networks, as discussed in Section~\ref{section:introduction}.
    Moreover, they either leverage traditional ML algorithm in a single layer framework (e.g., \cite{aswal2024advancing, gul2024improving}) or rely on highly complex models (e.g.,~\cite{yagiz2024transforming, mahdi2024advanced}). 
    As a result, these approaches may not perform effectively in highly mobile and distributed environments like IoV, which are constrained by limited computing resources and requires a scalable IDS solution.
    Therefore, this work proposed a scalable hierarchical IDS framework for IoV network that can mitigate the above challenges.

% =================================================
\section{The Proposed Hierarchical Classification Model}
\label{Proposed}
\begin{figure*}[!t]
   \centering
    \includegraphics[width = 0.75\linewidth]{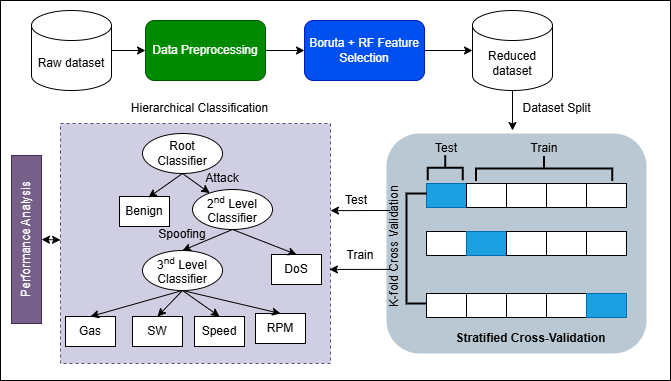}
    \caption{Hierarchical classification of the proposed IDS}
    \label{fig:model}
    \vspace{-15 pt}
\end{figure*}

As analyzed in Section~\ref{section:Literature_Review}, 
most existing studies focus on conventional flat and centralized multiclass classification approaches in IoV. 
    This raises concerns about the effectiveness of these models in accurately detecting network intrusions in distributed IoV network. 
    In addition, these models tend to produce a higher number of false negative cases (i.e., attacks are classified as normal), which is particularly concerning in practice. 
    To address this gap, our study aims to investigate hierarchical classification using a recent IoV dataset.  
    In this section, we present the design and architecture of our hierarchical classification approach, as depicted in Fig.~\ref{fig:model}. 
    Our framework begins with data preprocessing to clean and prepare the raw dataset. 
    Next, the Boruta algorithm, in combination with RF, is applied to select the features, retaining the most relevant features. 
    The resulting reduced dataset is then processed using a stratified cross-validation approach to ensure balanced class distribution across folds. 
    For each fold, a hierarchical classifier, as depicted in Fig.~\ref{fig:model}, is employed to perform multi-level classification.
    Here, we followed the three-level classifications of the dataset. 
    The first level identifies benign and attack categories. 
    The second level then has benign and specific attack types of attacks (i.e., DoS and Spoofing).
    Finally, the third level further identifies benign, DoS, and the subtypes of Spoofing attacks, resulting in the six categories in total.
    The remainder of this section will detail our proposed framework.
        
\subsection{Data Preprocessing}
In our proposed framework, the data preprocessing steps are as follows. 

\subsubsection{Dataset Description}
    The CIC-IoV2024 dataset is a comprehensive resource for studying IoV security, generated through experiments on a 2019 Ford car equipped with Electronic Control Units (ECUs). 
         It provides detailed insights into internal vehicle communication. 
         The dataset includes one primary feature corresponding to the ID and eight data features, each representing a byte of transmitted data. 
         For our experiments, we utilised the binary version, where each bit is represented as a binary feature. 
         % CICIoV2024 encompasses five attacks—DoS, gas spoofing, RPM spoofing, speed spoofing, and steering wheel spoofing—along with benign communication. 
         
    Table \ref{tab:dataset} shows the distribution of six classes (i.e., Benign, DoS, gas spoofing, RPM spoofing, speed spoofing, and steering wheel spoofing) in the dataset \cite{neto2024ciciov2024}. 
         Over 85\% of the records are normal, while the remaining records represent various types of attacks. 
         Gas records have the smallest contribution, accounting for less than 0.81\% of the total records.
         To address this unbalanced data issue, this work adopts stratified cross-validation that can divide the dataset into folds while maintaining the proportional distribution of each class within each fold.
    
\begin{table}
    \caption{Class distributions in the dataset}
    \label{tab:dataset}
    \centering
    
    \begin{tabular}{|l | l | r | r |}
        \hline
        \multicolumn{2}{|l|}{Class} & Records & Percentage \\
        \hline
        \multicolumn{2}{|l|}{Benign} & 1048575 & 85.04 \\
        \hline
        \multicolumn{2}{|l|}{DoS} & 74663 & 6.06 \\
        \hline
        
        \multirow{4}{*}{\begin{sideways}Spoofing \end{sideways}}& Gas & 9991 & 0.81 \\
        \cline{2-4}
        &RPM & 54900 & 4.45 \\
        \cline{2-4}
        &Speed & 24951 & 2.02 \\
        \cline{2-4}
        &Steering Wheel (SW) & 19977 & 1.62 \\
        \hline
        \multicolumn{2}{|l|}{Total} & 1233057
        
          & 100 \\
        \hline
    
    \end{tabular}
\end{table}
\subsubsection{Data Scaling}
    To eliminate the impact of  different feature's value in the datasets on the performance of machine learning algorithms, this study applies Min-Max normalisation. 
        This approach scales each feature's values to a range between 0 and 1. 
        The min-max normalisation value is calculated by~\ref{eq:min-max}.
        \begin{equation}
            \label{eq:min-max}
            X_{norm} = \frac{X - X_{min}}{X_{max} - X_{min}},
        \end{equation}
        where $X_{norm}$ is the normalised value of $X$, $X_{min}$ is the minimum value of $X$, and $X_{max}$ is the maximum value of $X$.
    
\subsubsection{Feature Selection}
    \textcolor{black}{
    Among feature selection methods in the literature, Boruta is known as an effective strategy for selecting robust features against noisy datasets and can deal with missing values and outliers. 
    Specifically, Boruta first generates shadow features by randomly shuffling the original features and then incorporating these features into the original dataset. 
    Then, by utilising RF, Boruta can reliably evaluate each feature's significance, even in the presence of noise or missing data. Farhana et al.\cite{farhana2023evaluation} demonstrated the effectiveness of the combination of Boruta and RF in feature selection to enhance the accuracy of detecting cyber attacks such as DDoS. They achieved a near perfect accuracy on CICIDS2017 dataset. 
        Therefore, we propose to create an ensemble feature selection approach by merging Boruta with RF.}
    Here, Boruta facilitates selecting the most relevant features, while Random Forest ranks the features based on their significance. 
    It is worth noting that Boruta is a distribution-free feature selection approach, unlike other traditional wrapper selection mechanisms, and computationally efficient, non-parametric, and can establish non-linear relationships between features and the target variable. 
    In addition, Boruta algorithm\cite{ahmed2021deep} is a wrapper-based feature selection method developed around the random forest algorithm.
    It identifies relevant features by comparing their importance to randomly generated ``shadow features.'' 
    The algorithm calculates the \textbf{Z-scores} of each feature's importance and evaluates their significance. 
    The main steps in the Boruta Algorithm are described below. 

            \begin{itemize}
                \item \textbf{Generate Shadow Features}:  
               Duplicate the original feature set, \( X \), and shuffle the values to create ``shadow features" (\( X_{\text{shadow}} \)) that are uncorrelated with the target variable \( y \). The dataset becomes:
               $[X, X_{\text{shadow}}].$
                
            \item \textbf{Compute Feature Importance}:  
               Using the random forest algorithm, calculate the importance of each feature \( X_i \) and its corresponding shadow feature \( X_{\text{shadow}, i} \) across all trees \( T \) in the forest by~\eqref{eq:importance}.
               \begin{equation}
                    \label{eq:importance}
                   \text{Importance}(X_i) = \frac{1}{T} \sum_{t=1}^{T} \mathbb{I}\{\text{OOB}_{t,\text{orig}} - \text{OOB}_{t,\text{perm}}\},
               \end{equation}

               where:
               \begin{itemize}
                   \item \( \mathbb{I} \) is the indicator function,
                   \item \( \text{OOB}_{t,\text{orig}} \) is the out-of-bag error without permutation,
                   \item \( \text{OOB}_{t,\text{perm}} \) is the error after permutation.
               \end{itemize}
            
            \item  \textbf{Calculate Z-Scores}:  
               For each feature \( X_i \), compute the Z-score by
               \begin{equation}
                    Z_i = \frac{\text{Importance}(X_i)}{\text{STD}(\text{Importance}(X_i))}, 
               \end{equation}
               
               where \( \text{STD} \) is the standard deviation of the feature importance values.
            
            \item  \textbf{Compare with Shadow Features}:  
               Determine the Maximum Z-score Among Shadow features (\( \text{MZSA} \)) and then classify features as:
               \begin{itemize}
                   \item \textbf{Confirmed}: \( Z_i > \text{MZSA} \)
                   \item \textbf{Unimportant}: \( Z_i < \text{MZSA} \)
                   \item \textbf{Tentative}: Requires further evaluation.
               \end{itemize}
            
            \item \textbf{Iterative Refinement}:  
               Shuffle and regenerate shadow features iteratively. 
               In each iteration, features are re-evaluated in until all features are classified as Confirmed or Unimportant, or the maximum iteration threshold (\( \text{maxRuns} \)) is reached.
            
            \item  \textbf{Final Selection}:  
               Features classified as Confirmed are selected as the most relevant predictors.
            
                The Boruta algorithm stops when all features are either Confirmed or Unimportant. 
                Note that features classified as Tentative after reaching \( \text{maxRuns} \) are finalized by comparing their median Z-scores with the best shadow feature's median Z-score.
            
       \end{itemize}

\subsubsection{Dataset Partition}
    Stratified cross-validation is used to divide the dataset into folds while maintaining the proportional distribution of each class within each fold. 
        This method provides a more reliable estimate of model performance, especially when dealing with imbalanced datasets where one class has significantly more samples than others. 
        In this study, both flat and hierarchical models were trained and evaluated using stratified 10-fold cross-validation. 
        The dataset was split into ten folds of equal size, with each fold preserving the class proportions. 
        This process was repeated for all ten folds, and the average classification performance was reported. 
        By utilising stratified cross-validation, both models achieve more accurate and dependable performance evaluations.
        
\subsubsection{Multi-level Classification Framework}
    After raw data is processed, a multi-level classification is employed to classify data into hierarchical classes: 
    \begin{itemize}
        \item \textit{Root Classifier (Level-1)}: Classify data into Benign or Attack,  
        \item \textit{Level-2 Classifier}: Further classify attack instances to identify the attack type, such as Spoofing or DoS (Denial of Service),
        \item \textit{Level-3 Classifier}: A deeper classification is performed to categorise spoofing attacks into specific types, including additional subtypes of spoofing, i.e., Gas, Steering Wheel (SW), Speed, RPM (Revolutions Per Minute).
    \end{itemize}
    The next section will discuss the distribution of our proposed multi-level classification framework in IoV.

\subsection{Distributing Proposed Hierarchical IDS Framework in IoV}
%%%%%%%%%%%%%%%%%%%%%%%%%%%% Figure of IoV network which will depict the deployment strategy is needed 

\begin{figure}[!t]
   \centering
    \includegraphics[width = \linewidth]{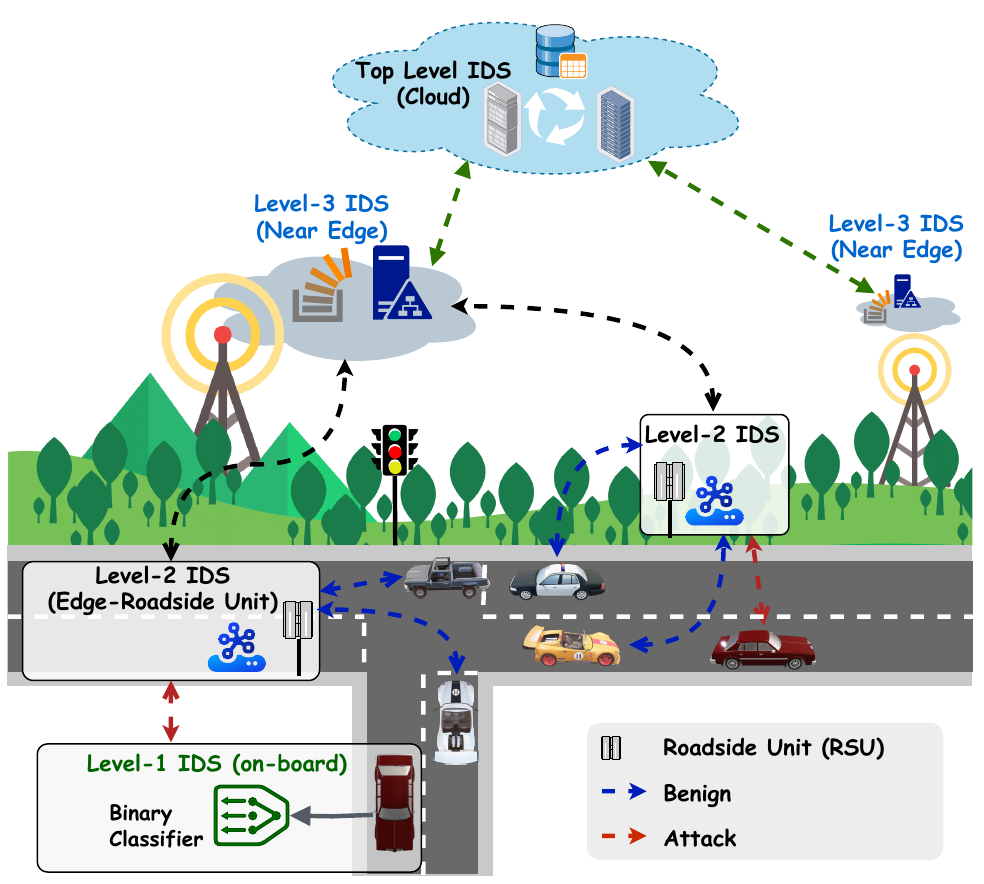}
    \caption{Hierarchical IDS for IoV.}
    \label{fig:system_model}
    \vspace{-15 pt}
\end{figure}

% {\color{red}This work considers an IoV network that is a layered, tier-based system composed of IoT-enabled vehicles, Edge-based Roadside Units (RSUs), Near Edge Nodes, and Cloud, as illustrated in Fig.~\ref{fig:system_model}. 
%     A highly significant advantage of hierarchical classification lies in its ability to distribute the different classifiers across various layers of the IoV network. 
%     This approach can take the advantage of the network's distributed architecture to effectively optimise resource utilisation and scalability by distributing classification responsibilities across the network's layers based on their specific capabilities. }
{\color{black}
    This work considers a typical IoV network facilitating diverse types of IoT applications, similar to those in~\cite{ang2019Deployment}, \cite{ kaiwartya2016Internet}, and \cite{duan2020Emerging}.
        Specifically, IoV is a layered, tier-based system composed of IoT-enabled vehicles, Edge-based Roadside Units (RSUs), Near Edge Nodes, and Cloud, as illustrated in Fig.~\ref{fig:system_model}. 
        Based on the layered architecture, we proposed a hierarchical IDS framework by distributing its components (i.e., classifiers) across different layers in the IoV system.      
        A significant advantage of hierarchical classification lies in its ability to distribute the different classifiers across various layers of the IoV network. 
        This approach can take the advantage of the network's distributed architecture to effectively optimise resource utilisation and scalability by distributing classification responsibilities across the network's layers based on their specific capabilities.
        Furthermore, by using this popular architecture, our proposed solution is highly applicable to existing IoV systems, where diverse types of IoV applications can be deployed.
    
    It is worth noting that the classifiers leveraged in our proposed hierarchical IDS framework, including Random Forest (RF) and Linear Regression (LR), are well-established and can work effectively in diverse scenarios~\cite{goodfellow2016Deep}, ranging from predicting fuel consumption and vehicle speed~\cite{pavlovic2020understanding},~\cite{yao2020vehicle} to anomaly detection in vehicle sensor data \cite{zuniga2019A}.
        Therefore, our proposed hierarchical IDS framework is neither specifically designed for a particular dataset, e.g., CIC-IoV2024, nor depends on a specific type of classifier.
        In fact, it can operate effectively in any related IoV security datasets and future effective classifiers.
        Recently, the Large Language Models (LLMs), often relying on transformer architectures, have been extensively explored in various areas, from text/image generation and code assisting to address challenges in wireless communications. 
        However, since LLMs were initially developed for natural language processing,  applying them directly to domain-specific tasks like intrusion detection in IoV systems can be challenging and would require extensive fine-tuning to achieve optimal performance~\cite{adjewa2024llm}. 
        Additionally, their sizes are usually large (e.g., 28 GB for LLAMA2-7B~\cite{llama2website}), placing an extreme burden in storage capabilities at the edge.
        Moreover, the inference process of LLM model like LLAMA2 can indeed take seconds, even with high-end GPUs~\cite{aman2023llama}.
        Thus, deploying such large models in resource-constrained environments like the IoV is currently unviable since it poses significant challenges in terms of storage capacity, computational power, and energy consumption.
    }   
    
% The deployment process is depicted in Figure \ref{fig:system_model}. 
{\color{black}In our approach, we propose deploying a well-trained root classifier of the hierarchical model within each IoV vehicle. 
    This root classifier can classify real-time IoT data as normal or attack, enabling vehicles to respond promptly to cyber threats without delays caused by communication overhead. 
    If a network traffic is regconized as attack, it will be then forwarded to the edge-based road side unit (where the second-level classifiers are deployed) via Vehicle-to-Infrastructure (V2I) Communications, e.g., Dedicated Short-Range Communications (DSRC), Cellular Vehicle-to-Everything (C-V2X), and Wireless Access in Vehicular Environments (WAVE)~\cite{ji2020survey}.
        
The second-level classifier, which identifies broad attack categories, is suggested to be deployed in the edge-based roadside units (RSUs). 
    These units are capable of receiving and processing classification requests from multiple vehicles to perform further analysis and categorisation of potential attacks.
    Near edge nodes are assigned the responsibility of detecting specific attack subcategories and then reporting them to the Cloud server. 
    This decentralised approach can to leverage the proximity of near edge nodes within specific areas, enabling quick and specific detection to emerging attacks within localised environments. 
    Thus, this targeted subcategory classification at near-edge nodes improves the overall efficiency and granularity of attack detection by the IDS in a distributed system setting.
    To obtain further insights on the identified attack traffic, the attack is then forwarded to the top-level classifier in our proposed hierarchical ID framework, usually via a high-bandwidth backhaul link~\cite{liu2013novel}.
 
The top level in our framework is cloud, which is tasked with critical responsibility of the hierarchical training of models and their components, ensuring robust and well-trained classifiers development. 
    After the training phase, these classifiers are strategically deployed at different layers within the IoV network to achieve not only attack detection granularity but also a high scalability for IDS systems. 
    In addition, this classifier distribution also plays a pivotal role of enhancing the capability of hierarchical model for detecting emerging attacks based on feedback from components in the lower layers.}

In this way, our decentralised approach can take advantage of IoV architectures by leveraging (i) the power of connected vehicles for binary classification, (ii) the proximity of edge nodes to data sources for quick and further analysis, (iii) the proximity of near edge nodes within specific localised environments for quick and specific detection to emerging attacks within these environments, and (iv) Cloud for training the hierarchical IDS and improved its capabilities with reporting information from other nodes in the IoT network.
    In contrast, a centralised flat classification model cannot be partitioned into distinct components for deployment at different levels of the IoV network. 
    Although a hierarchical model involves more classifiers and has a higher time complexity compared to a flat model, it offers significant advantages. 
    These include improved scalability, reduced communication overhead, and support for collaborative learning in a distributed, dynamic environment like IoV. 
    Thus, this architecture makes the hierarchical model a better fit for managing the complexities of IoV networks.

%=================================
\section{Results and Discussion}
\label{section:RESULTS AND DISCUSSION}
This section present our experimental setup and performance metrics. 
    Then, we evaluate the effectiveness of the hierarchical multi-classification model for IoV in terms of accuracy, precision, recall and F1-score.  
    Finally, we discuss our findings on the feature importance.

\subsection{Experimental Setup and Implementation}

Our experimental study was performed on a workstation equipped with AMD EPYC 7402P CPU (24 cores/48 threads), 256Gb DDR4 RAM, 3x Nvidia A100 GPUs (40Gb HBM2 Memory). 
    Our framework is deployed using Python 3.9 and rely on Pandas and NumPy libraries for data preprocessing. 
    Additionally, the widely recognised Scikit-learn toolkit is also used to leverage its wide range of algorithms, including popular machine learning algorithms (e.g., LR and RF), and effective accuracy and precision estimation metrics~\cite{pedregosa2011scikit}.

\textcolor{black}{
    The dataset CIC-IoV2024 is used for performance evaluation. 
        It consists of five attacks types, i.e., DoS, RPM, SPEED, STEERING\_WHEEL, and GAS.
        Note that our proposed hierarchical IDS framework does not depend on any specific dataset.
        In fact, our propose framework can detect as many attack types as those in the available training datasets.
        As such, if a new dataset is collected, the classifier at each layer can be retrained/fine-tuned to effectively detect any new attack types. 
        Thanks to the hierarchical architecture, these retraining/fine-tuning processes are performed in parallel, significantly reducing the time required to adapt to new threats.}
    
\textcolor{black}{In our experiment, we trained four machine learning algorithms: Decision Tree (DT), Random Forest (RF), Extreme Tree (ET), and Logistic Regression (LR), along with one deep learning algorithm, Deep Neural Network (DNN), in a federated learning setting. 
    However, since the tree-based algorithms (DT, RF, ET) showed nearly identical performance, we focused on presenting the performance of RF from the tree-based algorithms, along with LR and DNN in the federated setting. Additionally, we demonstrated the training and testing times of all the machine learning algorithms in both hierarchical and flat classification settings.}

% R4 - Comment 6

\subsection{Performance Metrics}

% \begin{figure}[!t]
%    \centering
%     \includegraphics[width = 0.99\linewidth]{IoV_Figure/featureSelection.png}
%     \caption{Feature selection approach}
%     \label{fig:featureselection}
%     \vspace{-10 pt}
% \end{figure} 

In this study, we used accuracy, precision, recall, and F1-score that are essential for assessing the performance of an IDS model. 
    Note that their significance can vary depending on the system's specific objectives and requirements. 
    Accuracy quantifies the proportion of accurate classifications made by the IDS. 
    However, relying solely on accuracy may not suitable performance metric for IDS, as this might not accurately reflect the system's capability to identify attacks belonging to a minority class within the dataset. 
    The reason is that precision refers to the proportion of genuine positive detection out of all positive detection. 
    Thus, high precision is essential in IDS in order to minimise false negatives (normal incorrectly detected as attack), which can result in false alarms. 
    On the other hand, recall measures the system's ability to reliably identify all instances of a particular class of attack. 
    Therefore, low recall suggests that the system is missing many attacks, which can pose a significant security risk.
    The combination of precision and recall yields the F1-score that quantifies the proportion of true positive identification relative to the total number of positive instances in the dataset. 
    F1-score is a valuable metric for IDS because this considers both false positives and false negatives and provides a balanced score between precision and recall.

The accuracy, precision, recall and F1-score are calculated by~\eqref{eq:accuracy}~\eqref{eq:precision}~\eqref{eq:recall}~\eqref{eq:f1-score}, respectively.

\begin{equation}
    \label{eq:accuracy}
    \text{Accuracy} = \frac{\text{TP + TN}}{\text{TP + TN + FP + FN}},
\end{equation}

\begin{equation}
    \label{eq:precision}
   \text{Precision} = \frac{\text{TP}}{\text{TP+FP}},
\end{equation}

\begin{equation}
    \label{eq:recall}
    \text{Recall} = \frac{\text{TP}}{\text{TP+FN}},
\end{equation}

\begin{equation}
    \label{eq:f1-score}
    \text{F1-score} = 2\times \frac{\text{Precision $\times$ Recall}}{\text{Precision + Recall}},
\end{equation}

where TP = true positive, TN = true negative, FP = false positive, and FN = false negative.

In our experiments, we used Stratified cross-validation which can effectively addresses the imbalance in test datasets by maintaining balanced class distributions across folds, ensuring accurate evaluation metrics such as precision, recall, and F1-score. 
    This approach guards against biased evaluations driven by dominant classes and enhances model robustness to dataset variability. Overall, it provides reliable estimates of generalization performance for models trained on imbalanced data, crucial for real-world applications.

\subsection{Impact of Feature Selection on Model Accuracy}

\begin{figure}
		\centering
		$\begin{array}{c}
			\includegraphics[width=1\linewidth]{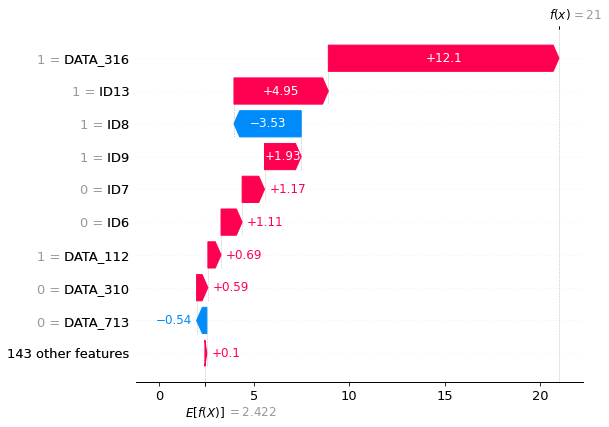} \\
			\text{(a) SHAP-XGBoost}  \\
			\includegraphics[width=1\linewidth]{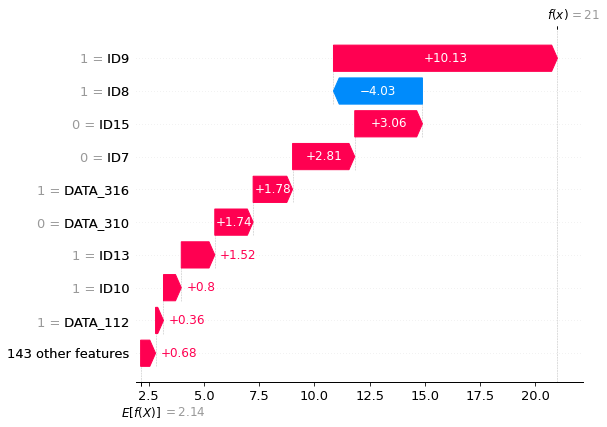} \\
			\text{(b) SHAP-RF}
		\end{array}$
		\caption{Feature importance of CIC-IoV2024 datasets.}
		\label{fig:featureimportance}
        \vspace{-15 pt}
\end{figure}

\begin{figure}[!thbp]
   \centering
    \includegraphics[width = 0.99\linewidth]{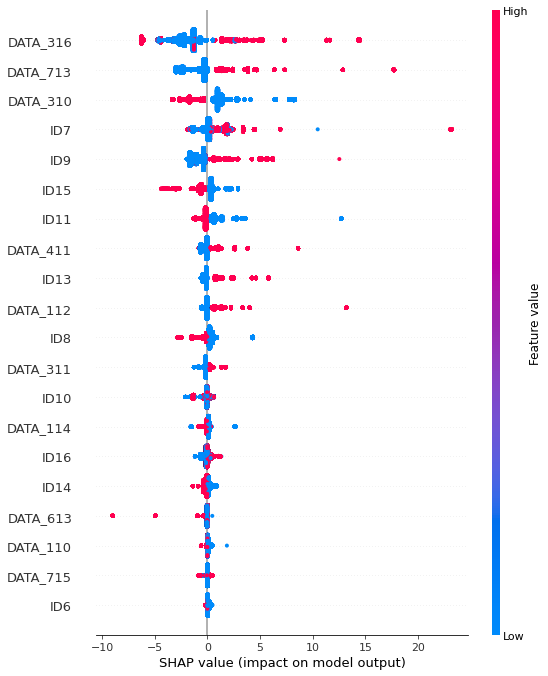}
    \caption{Feature importance using SHAP-RF on the outcome}
    \label{fig:shapfeatureoutcome}
\end{figure} 

\begin{figure}[!t]
   \centering
    \includegraphics[width = \linewidth]{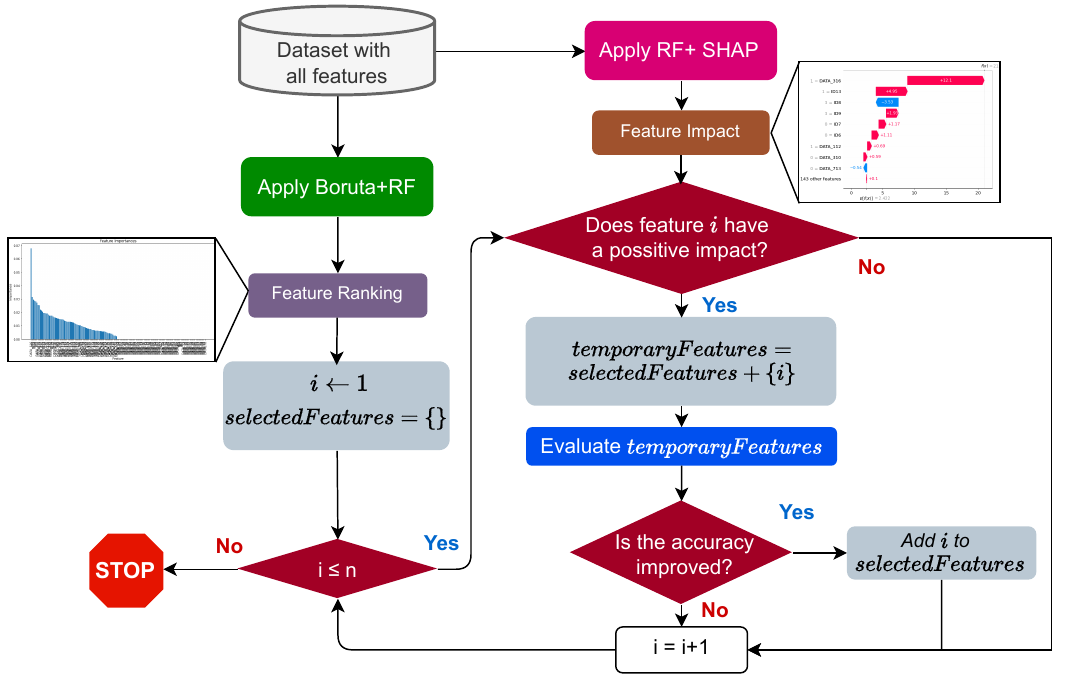}
    \caption{Flowchart of feature selection approach}
    \label{fig:flowchartfeatureselection}
    \vspace{-20 pt}
\end{figure}

Our feature selection approach is presented in Fig.~\ref{fig:flowchartfeatureselection}. 
    In particular, we utilised two techniques to select the most significant features for our model: Boruta with RF (namely Boruta+RF) and SHAP (SHapley Additive exPlanations) with RF, i.e., SHAP-RF. 
    First, Boruta+RF was employed to rank the features based on their importance. 
    Subsequently, SHAP-RF was used to analyze the impact and importance of each feature on the classification of specific attacks.
    Interestingly, our results revealed that:

    \begin{itemize}
        \item At the root level (Benign vs. Attack) and second-level classifier (Spoofing vs. DoS), a minimum of 11 features (equivalent to 7\%) from the ranked feature list was sufficient to achieve 100\% accuracy. 
        % Note that the total number of feature is {\color{red}152}. 
        \item For the third level (GAS, SW, Speed, and RPM classification), 18 features were required to achieve 100\% accuracy. 
            Here, we need to include the $45^{th}$ feature with first seventeen top features. 
            We can skip features $18^{th}$ to $44^{th}$ from the ranked feature list. 
            Our systematic approach involves ranking all features; however, through trial and error combined with SHAP analysis, we observed that features $18^{th}$ to $44^{th}$ contribute minimally to the outcome. As a result, we prioritized feature $45^{th}$ while skipping features $18^{th}$ to $44^{th}$, as their positive impact on the outcome was determined to be negligible. This streamlined selection process ensures that we focus on the most impactful features, enhancing the model's efficiency and performance.
    \end{itemize}

On the other hand, when relying solely on Boruta+RF, at least 45 features from the ranked feature list were necessary to attain 100\% accuracy for the third-level classifier. 
    Thus, these findings highlight the significance of combining SHAP and Boruta with RF, as this integrated approach not only reduces the number of required features but also enhances interpretability and classification efficiency. 
    The feature importance using SHAP with XGBoost and RF are presented in Figs.~\ref{fig:featureimportance} and \ref{fig:shapfeatureoutcome}. 
    Specifically, Fig.~\ref{fig:featureimportance} illustrates the importance of various features in determining the outcome ``benign.'' 
    The analysis reveals that the feature ID8 has a negative impact on the outcome, indicating that it should be excluded from the selected features. Conversely, features such as DATA\_316, ID13, ID9, and several others positively contribute to the outcome and are valuable for inclusion in the feature set.

\begin{figure}[t]
		\centering
		$\begin{array}{c}
			\includegraphics[width=0.9\linewidth]{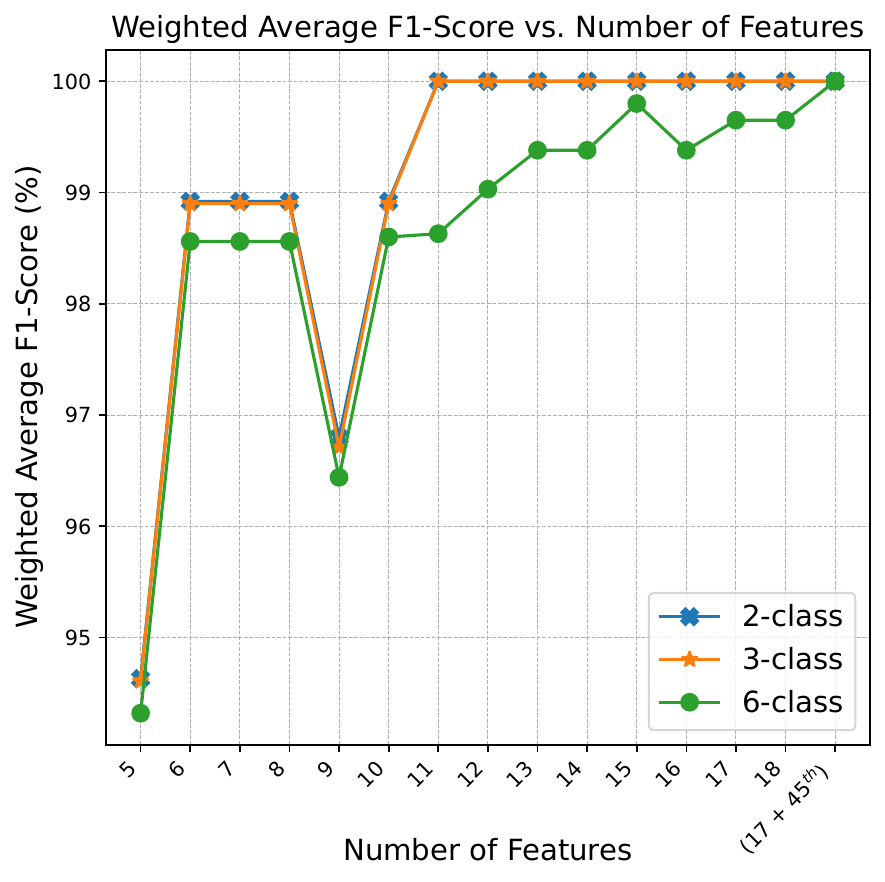} \\
			\text{(a)}  \\
			\includegraphics[width=0.9\linewidth]{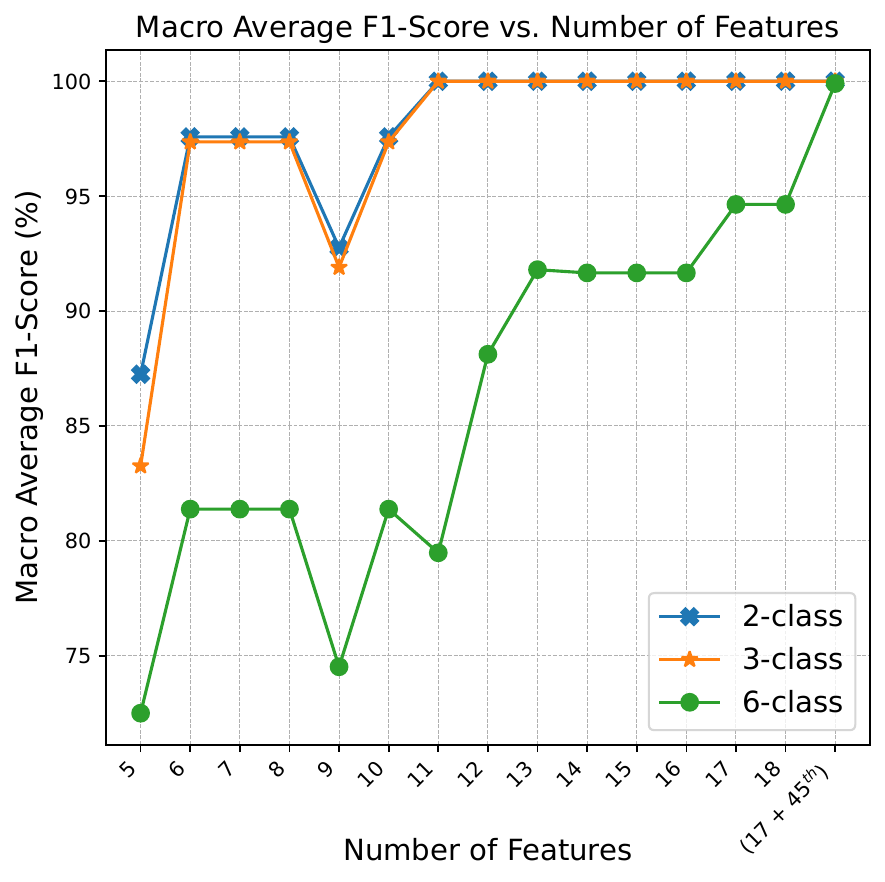} \\
			\text{(b)}
		\end{array}$
		\caption{(a) Weighted Average F1-Score vs. Number of Features for Different Classes and (b) Macro Average F1-Score vs. Number of Features for Different Classes.}
		\label{fig:f1scorefeatureselection}
		\vspace{-15 pt}
\end{figure}

Next, we investigated how the weighted average F1-score varies with the number of features for 2-class, 3-class, and 6-class classification tasks, as shown in~Fig.~\ref{fig:f1scorefeatureselection}(a).
    Note that 2-class, 3-class, and 6-class corresponds to the benign-attack, benign-DoS-Spoofing, and  benign-DoS-Spoofing subtypes classification tasks, respectively.    
    For 2-class and 3-class tasks, the F1-score reaches a perfect 100\% with only 11 features, showing that fewer features are sufficient for simpler classification tasks. 
    In contrast, the 6-class scenario achieves a maximum F1-score of 99.65\% with 17 or 18 top-ranked features, reflecting the greater complexity of multi-class classification. 
    A noticeable drop in performance across all scenarios at 9–10 features suggests the introduction of noise or irrelevant features. 
    The most important insight is that while adding features generally improves performance, the benefit plateaus once sufficient information is available (e.g., 11 features for 2-class and 3-class problems). 
    Thus, for complex tasks like 6-class classification, feature relevance and careful selection are crucial to maximising performance.
Additionally, Fig.~\ref{fig:f1scorefeatureselection}(a) shows that the Weighted Average F1-Score for Random Forest with the corrected feature set including ``17 Features + $45^{th}$ Feature.'' 
    This highlights how performance improves with additional features, achieving perfect scores for all classes when the $45^{th}$ feature is combined with the top 17 features.
  
The graph in Fig.~\ref{fig:f1scorefeatureselection}(b) shows how the Macro Average F1-score (\%) varies with the number of features for 2-class, 3-class, and 6-class classification tasks. For 2-class and 3-class problems, the F1-score improves steadily, reaching 100\% with 12 features and remaining consistent thereafter. In contrast, the 6-class scenario shows lower performance, starting at 72.5\% with 5 features and gradually increasing to 94.64\% with 17–18 features. A noticeable dip in F1-scores is observed at 9–10 features across all scenarios, likely due to noise or irrelevant features. Overall, the graph highlights that more features improve performance, but the benefit plateaus for simpler tasks.

The graph illustrates that classifying specific attack categories (6 classes) is more challenging for machine learning algorithms and requires more features to accurately classify these categories. While perfect F1-scores can be achieved with fewer features for 2-class and 3-class problems, the algorithm struggles with the complexity of 6-class attack classification. Using Boruta and RF feature selection techniques, it was determined that at least 45 features are needed for third-level classification. However, SHAP analysis reveals that adding the 45th ranked feature from Boruta results to a subset of 17 features enables the model to achieve perfect F1-scores and accuracy for all attack categories at the third level. This underscores the importance of combining SHAP and Boruta feature selection methods. It is also noted that most of the existing works on CIC-IoV2024 datasets have not explored third-level attack classification.

\subsection{Hierarchical Model Performance Across Levels}

\begin{table*}[t]
    \caption{Performance of Root-Level and Second-Level Classifiers}
    \centering
    \label{tab:root2ndlevelperformance}
    \begin{tabular}{|c|l|llllll|}
        \hline
        \multirow{3}{*}{\textbf{Model}} & \multicolumn{1}{c|}{\multirow{3}{*}{\textbf{Category (2-class)}}} & \multicolumn{6}{c|}{\textbf{Root Level Classifier}} \\ \cline{3-8} 
         & \multicolumn{1}{c|}{} & \multicolumn{3}{c|}{\textbf{10 Features}} & \multicolumn{3}{c|}{\textbf{11 Features}} \\ \cline{3-8} 
         & \multicolumn{1}{c|}{} & \multicolumn{1}{l|}{\textbf{Precision}} & \multicolumn{1}{l|}{\textbf{Recall}} & \multicolumn{1}{l|}{\textbf{F1-score}} & \multicolumn{1}{l|}{\textbf{Precision}} & \multicolumn{1}{l|}{\textbf{Recall}} & \textbf{F1-score} \\ \hline
        \multirow{4}{*}{RF} & BENIGN & \multicolumn{1}{l|}{98.79} & \multicolumn{1}{l|}{100} & \multicolumn{1}{l|}{99.39} & \multicolumn{1}{l|}{100} & \multicolumn{1}{l|}{100} & 100 \\ \cline{2-8} 
         & ATTACK & \multicolumn{1}{l|}{100} & \multicolumn{1}{l|}{91.88} & \multicolumn{1}{l|}{95.77} & \multicolumn{1}{l|}{100} & \multicolumn{1}{l|}{100} & 100 \\ \cline{2-8} 
         & macro avg & \multicolumn{1}{l|}{99.4} & \multicolumn{1}{l|}{95.94} & \multicolumn{1}{l|}{97.58} & \multicolumn{1}{l|}{100} & \multicolumn{1}{l|}{100} & 100 \\ \cline{2-8} 
         & weighted avg & \multicolumn{1}{l|}{98.95} & \multicolumn{1}{l|}{98.94} & \multicolumn{1}{l|}{98.92} & \multicolumn{1}{l|}{100} & \multicolumn{1}{l|}{100} & 100 \\ \hline
        \multirow{4}{*}{LR} & BENIGN & \multicolumn{1}{l|}{97.99} & \multicolumn{1}{l|}{99.44} & \multicolumn{1}{l|}{98.71} & \multicolumn{1}{l|}{100} & \multicolumn{1}{l|}{99.93} & 99.96 \\ \cline{2-8} 
         & ATTACK & \multicolumn{1}{l|}{95.89} & \multicolumn{1}{l|}{86.48} & \multicolumn{1}{l|}{90.94} & \multicolumn{1}{l|}{99.53} & \multicolumn{1}{l|}{100} & 99.77 \\ \cline{2-8} 
         & macro avg & \multicolumn{1}{l|}{96.94} & \multicolumn{1}{l|}{92.96} & \multicolumn{1}{l|}{94.83} & \multicolumn{1}{l|}{99.77} & \multicolumn{1}{l|}{99.96} & 99.87 \\ \cline{2-8} 
         & weighted avg & \multicolumn{1}{l|}{97.72} & \multicolumn{1}{l|}{97.74} & \multicolumn{1}{l|}{97.69} & \multicolumn{1}{l|}{99.94} & \multicolumn{1}{l|}{99.94} & 99.94 \\ \hline
        \multicolumn{1}{|l|}{} & \textbf{Category (3-class)} & \multicolumn{6}{c|}{\textbf{2nd Level Classifier}} \\ \hline
        \multirow{5}{*}{RF} & BENIGN & \multicolumn{1}{l|}{98.79} & \multicolumn{1}{l|}{100} & \multicolumn{1}{l|}{99.39} & \multicolumn{1}{l|}{100} & \multicolumn{1}{l|}{100} & 100 \\ \cline{2-8} 
         & SPOOFING & \multicolumn{1}{l|}{100} & \multicolumn{1}{l|}{86.35} & \multicolumn{1}{l|}{92.68} & \multicolumn{1}{l|}{100} & \multicolumn{1}{l|}{100} & 100 \\ \cline{2-8} 
         & DoS & \multicolumn{1}{l|}{100} & \multicolumn{1}{l|}{100} & \multicolumn{1}{l|}{100} & \multicolumn{1}{l|}{100} & \multicolumn{1}{l|}{100} & 100 \\ \cline{2-8} 
         & macro avg & \multicolumn{1}{l|}{99.6} & \multicolumn{1}{l|}{95.45} & \multicolumn{1}{l|}{97.36} & \multicolumn{1}{l|}{100} & \multicolumn{1}{l|}{100} & 100 \\ \cline{2-8} 
         & weighted avg & \multicolumn{1}{l|}{98.95} & \multicolumn{1}{l|}{98.94} & \multicolumn{1}{l|}{98.9} & \multicolumn{1}{l|}{100} & \multicolumn{1}{l|}{100} & 100 \\ \hline
        \multirow{5}{*}{LR} & BENIGN & \multicolumn{1}{l|}{97.99} & \multicolumn{1}{l|}{99.44} & \multicolumn{1}{l|}{98.71} & \multicolumn{1}{l|}{100} & \multicolumn{1}{l|}{99.93} & 99.96 \\ \cline{2-8} 
         & SPOOFING & \multicolumn{1}{l|}{94.89} & \multicolumn{1}{l|}{86.35} & \multicolumn{1}{l|}{90.42} & \multicolumn{1}{l|}{99.22} & \multicolumn{1}{l|}{100} & 99.61 \\ \cline{2-8} 
         & DoS & \multicolumn{1}{l|}{97.4} & \multicolumn{1}{l|}{86.69} & \multicolumn{1}{l|}{91.73} & \multicolumn{1}{l|}{100} & \multicolumn{1}{l|}{100} & 100 \\ \cline{2-8} 
         & macro avg & \multicolumn{1}{l|}{96.76} & \multicolumn{1}{l|}{90.82} & \multicolumn{1}{l|}{93.62} & \multicolumn{1}{l|}{99.74} & \multicolumn{1}{l|}{99.98} & 99.86 \\ \cline{2-8} 
         & weighted avg & \multicolumn{1}{l|}{97.72} & \multicolumn{1}{l|}{97.74} & \multicolumn{1}{l|}{97.69} & \multicolumn{1}{l|}{99.94} & \multicolumn{1}{l|}{99.94} & 99.94 \\ \hline
    \end{tabular}
\end{table*}

\begin{table*}[!h]
\caption{Performance of third level classifier}
\centering
\label{tab:3rdlevelperformance}
\begin{tabular}{|c|l|llllll|}
\hline
\multirow{3}{*}{\textbf{Model}} & \multicolumn{1}{c|}{\multirow{3}{*}{\textbf{Categories (6-class)}}} & \multicolumn{6}{c|}{\textbf{3rd Level Classifier}} \\ \cline{3-8} 
 & \multicolumn{1}{c|}{} & \multicolumn{3}{c|}{\textbf{17 Features}} & \multicolumn{3}{c|}{\textbf{18 Features}} \\ \cline{3-8} 
 & \multicolumn{1}{c|}{} & \multicolumn{1}{l|}{\textbf{Precision}} & \multicolumn{1}{l|}{\textbf{Recall}} & \multicolumn{1}{l|}{\textbf{F1-score}} & \multicolumn{1}{l|}{\textbf{Precision}} & \multicolumn{1}{l|}{\textbf{Recall}} & \textbf{F1-score} \\ \hline
\multirow{8}{*}{RF} & BENIGN & \multicolumn{1}{l|}{100} & \multicolumn{1}{l|}{100} & \multicolumn{1}{l|}{100} & \multicolumn{1}{l|}{100} & \multicolumn{1}{l|}{100} & 100 \\ \cline{2-8} 
 & DoS & \multicolumn{1}{l|}{100} & \multicolumn{1}{l|}{100} & \multicolumn{1}{l|}{100} & \multicolumn{1}{l|}{100} & \multicolumn{1}{l|}{100} & 100 \\ \cline{2-8} 
 & RPM & \multicolumn{1}{l|}{100} & \multicolumn{1}{l|}{99.99} & \multicolumn{1}{l|}{99.99} & \multicolumn{1}{l|}{100} & \multicolumn{1}{l|}{99.99} & 99.99 \\ \cline{2-8} 
 & SPEED & \multicolumn{1}{l|}{100} & \multicolumn{1}{l|}{80} & \multicolumn{1}{l|}{88.89} & \multicolumn{1}{l|}{100} & \multicolumn{1}{l|}{100} & 100 \\ \cline{2-8} 
 & STEERING\_WHEEL & \multicolumn{1}{l|}{100} & \multicolumn{1}{l|}{100} & \multicolumn{1}{l|}{100} & \multicolumn{1}{l|}{100} & \multicolumn{1}{l|}{100} & 100 \\ \cline{2-8} 
 & GAS & \multicolumn{1}{l|}{66.67} & \multicolumn{1}{l|}{100} & \multicolumn{1}{l|}{80} & \multicolumn{1}{l|}{99.94} & \multicolumn{1}{l|}{100} & 99.97 \\ \cline{2-8} 
 & macro avg & \multicolumn{1}{l|}{94.44} & \multicolumn{1}{l|}{96.67} & \multicolumn{1}{l|}{94.81} & \multicolumn{1}{l|}{99.99} & \multicolumn{1}{l|}{100} & 99.99 \\ \cline{2-8} 
 & weighted avg & \multicolumn{1}{l|}{99.76} & \multicolumn{1}{l|}{99.65} & \multicolumn{1}{l|}{99.66} & \multicolumn{1}{l|}{100} & \multicolumn{1}{l|}{100} & 100 \\ \hline
\multirow{8}{*}{LR} & BENIGN & \multicolumn{1}{l|}{{100}} & \multicolumn{1}{l|}{{100}} & \multicolumn{1}{l|}{{100}} & \multicolumn{1}{l|}{\textbf{100}} & \multicolumn{1}{l|}{{100}} & {100} \\ \cline{2-8} 
 & DoS & \multicolumn{1}{l|}{100} & \multicolumn{1}{l|}{100} & \multicolumn{1}{l|}{100} & \multicolumn{1}{l|}{100} & \multicolumn{1}{l|}{100} & 100 \\ \cline{2-8} 
 & RPM & \multicolumn{1}{l|}{100} & \multicolumn{1}{l|}{99.99} & \multicolumn{1}{l|}{99.99} & \multicolumn{1}{l|}{100} & \multicolumn{1}{l|}{99.99} & 99.99 \\ \cline{2-8} 
 & SPEED & \multicolumn{1}{l|}{100} & \multicolumn{1}{l|}{80} & \multicolumn{1}{l|}{88.89} & \multicolumn{1}{l|}{100} & \multicolumn{1}{l|}{100} & 100 \\ \cline{2-8} 
 & STEERING\_WHEEL & \multicolumn{1}{l|}{100} & \multicolumn{1}{l|}{99.96} & \multicolumn{1}{l|}{99.98} & \multicolumn{1}{l|}{100} & \multicolumn{1}{l|}{100} & 100 \\ \cline{2-8} 
 & GAS & \multicolumn{1}{l|}{66.67} & \multicolumn{1}{l|}{100} & \multicolumn{1}{l|}{80} & \multicolumn{1}{l|}{99.94} & \multicolumn{1}{l|}{100} & 99.97 \\ \cline{2-8} 
 & macro avg & \multicolumn{1}{l|}{94.44} & \multicolumn{1}{l|}{96.66} & \multicolumn{1}{l|}{94.81} & \multicolumn{1}{l|}{99.99} & \multicolumn{1}{l|}{100} & 99.99 \\ \cline{2-8} 
 & weighted avg & \multicolumn{1}{l|}{99.76} & \multicolumn{1}{l|}{99.64} & \multicolumn{1}{l|}{99.66} & \multicolumn{1}{l|}{100} & \multicolumn{1}{l|}{100} & 100 \\ \hline
\end{tabular}
\end{table*}

The results in Table \ref{tab:root2ndlevelperformance} highlight the class-wise performance of root-level (2-class) and second-level (3-class) classifiers using RF and LR models, demonstrating the impact of features on precision, recall, and F1-scores.

\subsubsection{Root-Level Classifier (2-Class)}

\begin{itemize}
    \item RF Model: With 11 features, the RF model achieves perfect precision, recall, and F1-scores (100\%) for both categories (BENIGN and ATTACK), compared to slightly lower performance with 10 features (macro F1 = 97.58\%).
    \item LR Model: Similarly, the LR model significantly improves with 11 features, achieving near-perfect macro F1-scores (99.87\%) compared to 94.83\% with 10 features. This is due to enhanced performance in distinguishing ATTACK and BENIGN categories.
    
\end{itemize}

\subsubsection{Second-Level Classifier (3-Class)}
\begin{itemize}
    \item RF Model: With 11 features, RF achieves perfect scores (100\%) for all three categories (BENIGN, SPOOFING, and DoS). At 10 features, macro F1-scores are slightly lower (97.36\%), primarily due to lower recall for the SPOOFING category (86.35\%).
    \item LR Model: The LR model improves significantly with 11 features, achieving macro F1-scores of 99.86\% compared to 93.62\% with 10 features. While precision and recall for SPOOFING and DoS categories are slightly lower with 10 features, they reach near-perfect levels with 11 features.
\end{itemize}

Adding one additional feature (from 10 to 11) leads to substantial improvements in both root-level and second-level classifications, particularly for challenging categories like SPOOFING. The RF model achieves perfect classification with 11 features for both root-level and second-level tasks, outperforming LR slightly in terms of consistency.
The results underscore the importance of feature selection in improving precision, recall, and F1-scores, especially for multi-class problems.

Table \ref{tab:3rdlevelperformance} demonstrates the performance of RF and LR models as 3rd-level classifiers for a 6-class problem, comparing results with 17 and 18 features. For both models, adding the 45th feature results in a significant improvement in performance, particularly for the challenging GAS and SPEED categories. For GAS, the F1-score increases from 80 to 99.97, while SPEED improves from 88.89 to 100. All other categories, such as BENIGN, DoS, RPM, and STEERING WHEEL (SW), already exhibit near-perfect scores with 17 features, which are maintained with 18 features.

The macro average F1-score increases from 94.81\% to 99.99\%, reflecting the balanced improvement across all classes with the additional feature. The weighted average F1-score also reaches a perfect 100\% with 18 features, showing the overall robustness of both models across imbalanced class distributions. Both RF and LR achieve identical results across all metrics with 18 features, highlighting their comparable efficacy when sufficient features are available. The GAS category demonstrates the most significant improvement, with its F1-score rising from 80\% to 99.97\% after adding the 18th feature, emphasizing the importance of including relevant features for specific class challenges.

    \begin{figure}[!]
       \centering
        \includegraphics[width = \linewidth]{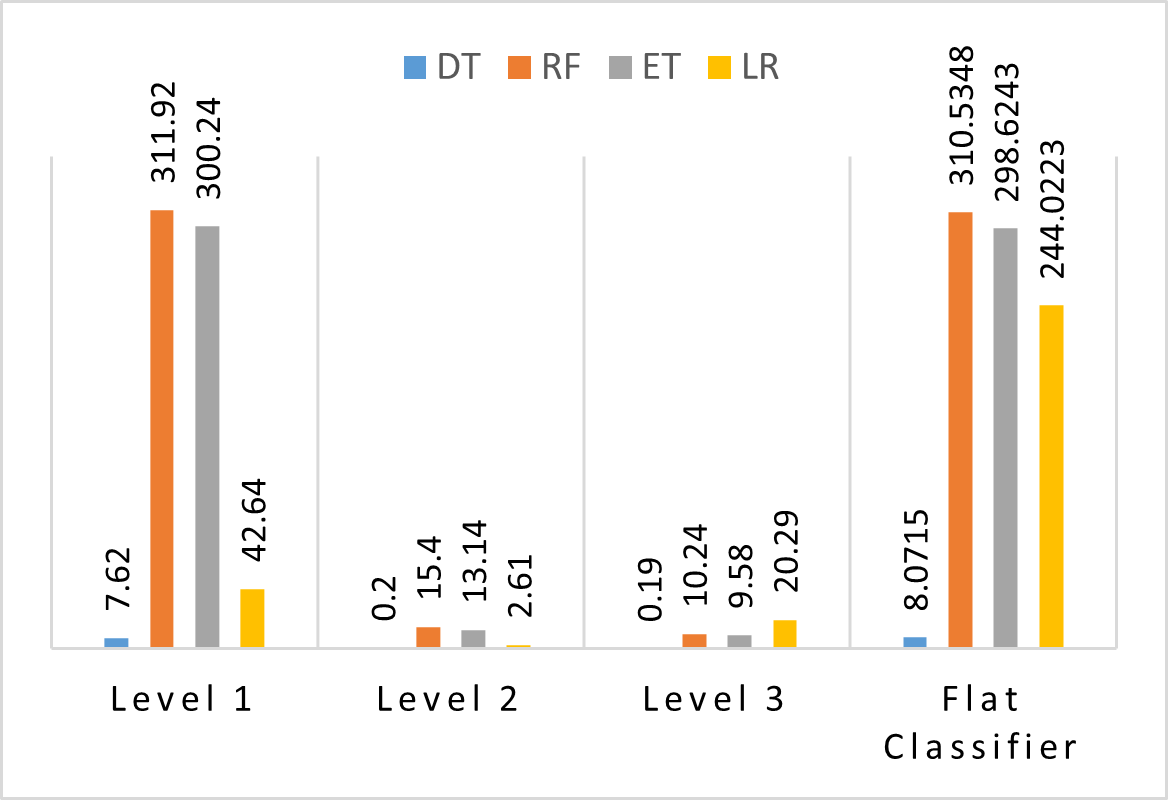}
        \caption{Level wise Training time for Hierarchical and Flat classifier}
        \label{fig:levelwisetime}
        \vspace{-5 pt}
    \end{figure} 

    \begin{figure}[!]
       \centering
        \includegraphics[width = \linewidth]{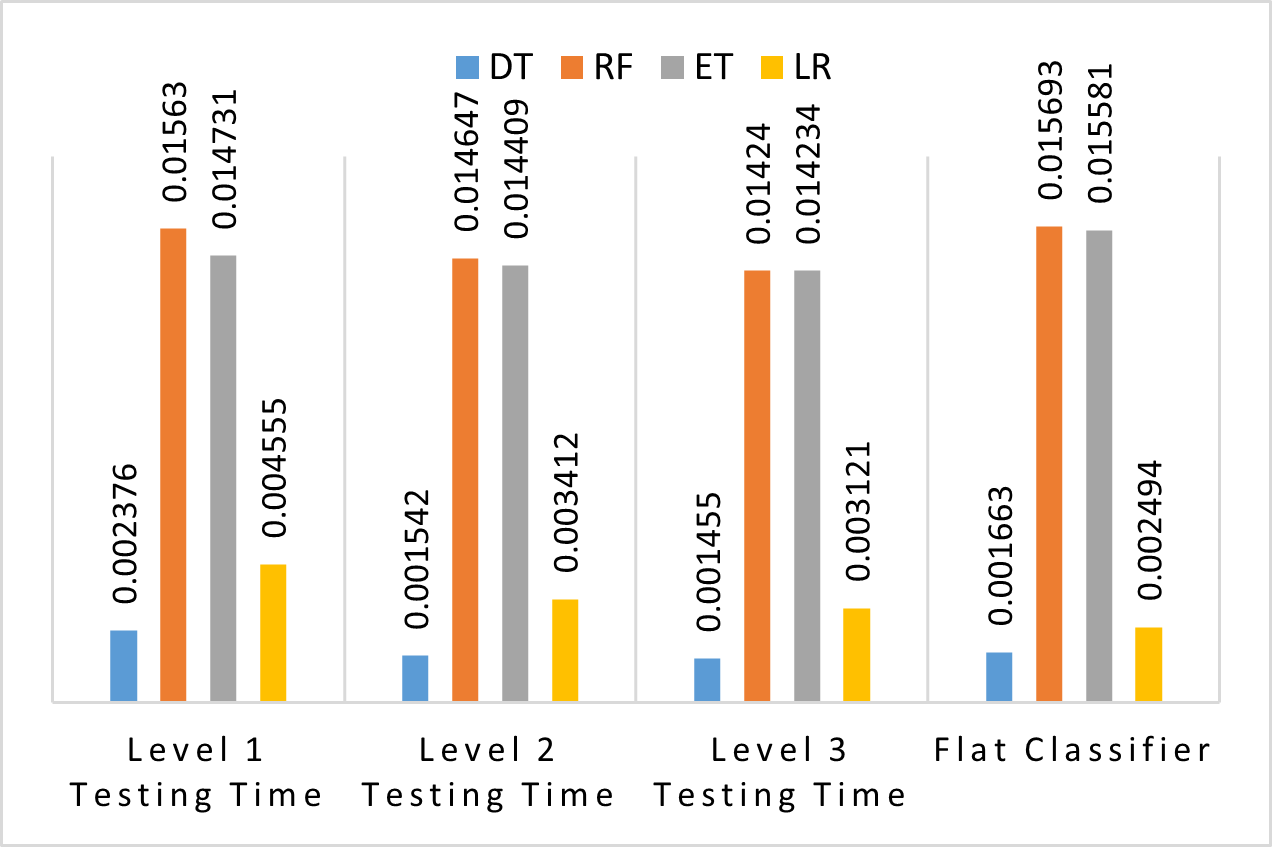}
        \caption{Level wise Testing time for single instance in Hierarchical and Flat classifier}
        \label{fig:singlelevelwisetime}
        \vspace{-10 pt}
    \end{figure} 

\subsubsection{Training and Testing Time}
\textcolor{black}{
Figures \ref{fig:levelwisetime} and \ref{fig:singlelevelwisetime} present the training and testing time of a single instance for the hierarchical and flat classifiers respectively. Figure \ref{fig:singlelevelwisetime} shows that major contributor to testing time is the choice of machine learning model. Random Forest (RF) and Extra Trees (ET), being ensemble models composed of multiple decision trees, require each input sample to traverse all trees for prediction, which becomes computationally intensive-especially when performed across three hierarchical levels. In contrast, Decision Tree (DT) models use a single tree for inference, resulting in faster predictions. Logistic Regression (LR) is even more efficient, involving only basic mathematical operations (e.g., dot products), making it the fastest model during inference, despite a slightly higher training time. However, as shown in Fig.~\ref{fig:singlelevelwisetime}, the testing time for a single instance is negligible. In addition, training is relatively lightweight for most models, making the hierarchical architecture practical for real-time applications.}

\subsection{Comparison of the Proposed Model with Existing Models}

The comparison presented in Table \ref{tab:modelcomparison} highlights the effectiveness of the proposed solution in achieving superior performance for 2-class, 3-class, and 6-class classification tasks on the CIC-IoV2024 dataset. Here, we compare our approach with existing solutions in the literature~\cite{neto2024ciciov2024, yagiz2024transforming,aswal2024advancing, mahdi2024advanced, gul2024improving}. In addition, Federated Learning (FL) has gained significant attention for deploying IDS in distributed environments due to its ability to enhance security and preserve privacy. % To ensure privacy and security, many researchers recommend using federated learning models. 
As such, for evaluation and comparison purposes, we implemented FL using the Flower framework. 
   % Our FL setup consists of 10 clients and a central server to facilitate the training and coordination process. 

{\color{black} To evaluate the performance of federated learning (FL) on the CIC-IoV2024 dataset, we employed the Flower framework to simulate a federated setting. 
    Specifically, we distributed the dataset across 10 clients and trained a Deep Neural Network (DNN) model over 5 communication rounds. 
    Each client performed local training of this DNN using its own subset of data. 
    Then, they sent their model parameters to a central server. 
    Upon receiving clients' model parameters, the central server aggregated these parameters using the Federated Averaging (FedAvg) algorithm~\cite{mcmahan2017communication}. 
    After that, the server send the model's parameters back to clients, and one communication round is completed. 
    The process is repeated until the learning converges.  We experimented with three feature configurations: 10, 11, and 18 features according to the classification task at each tier in our proposed approach. The classification tasks included binary, ternary, and six-class settings. The DNN architecture used in our FL experiments consisted of three hidden layers with ReLU activation, dropout regularization, and a softmax output layer. For training, we used categorical cross-entropy loss with the Adam optimizer.  The number of local epochs (50) and batch size (25) were fixed across all clients to ensure consistency.}

    The proposed hierarchical model achieves 100\% accuracy for 2-class and 3-class tasks with only 11 features and for 6-class tasks with 18 features, demonstrating its efficiency and suitability for resource-constrained IoV networks. 
    Similarly, FL also achieves very high accuracy. 
    However, conventional FL needs to maintain a full model at each devices, making it less effective due to the limited resources in IoV devices. 
    
{\color{black}In the literature, some existing methods like Neto et al.\cite{neto2024ciciov2024} achieve only 95\% accuracy, while others, such as Yagiz et al.\cite{yagiz2024transforming} and Aswal et al.\cite{aswal2024advancing}, claim high accuracy but have limitations.
    They either consider few attack categories or rely on all 152 features, which make them less practical for IoV networks. 
    In contrast, {our proposed model’s} ability to achieve perfect accuracy with minimal features underscores its potential for scalable and efficient IoV security solutions.}

\begin{table*}[!h]
    \caption{Performance Comparisons across 2-class, 3-class, and 6-class classification tasks in terms of accuracy.}
    \centering
    \label{tab:modelcomparison}
    \begin{tabular}{|l|l|l|l|l|}
        \hline
        \multicolumn{1}{|c|}{\textbf{Model}} & \multicolumn{1}{c|}{\textbf{\begin{tabular}[c]{@{}c@{}}2-Class \end{tabular}}} & \multicolumn{1}{c|}{\textbf{\begin{tabular}[c]{@{}c@{}}3-Class\end{tabular}}} & \multicolumn{1}{c|}{\textbf{\begin{tabular}[c]{@{}c@{}}6-Class \end{tabular}}} & \multicolumn{1}{c|}{\textbf{Remarks}} \\ \hline
        \begin{tabular}[c]{@{}l@{}}Proposed Model (10 Features)\end{tabular} & 98.94 & 98.94 & 98.94 & Default   parameters. \\ \hline
        \begin{tabular}[c]{@{}l@{}}Proposed Model (11 Features)\end{tabular} & 100 & 100 & 98.94 & Default   parameters. \\ \hline
        \begin{tabular}[c]{@{}l@{}}Proposed Model (18 Features)\end{tabular} & 100 & 100 & 100 & Default   parameters. \\ \hline
        \begin{tabular}[c]{@{}l@{}}Federated Learning (10 Features)\end{tabular} & 97.01 & 96.91 & 95.23 & \begin{tabular}[c]{@{}l@{}}Flower Framework 5 rounds with 10 clients.\end{tabular} \\ \hline
        \begin{tabular}[c]{@{}l@{}}Federated Learning (11 Features)\end{tabular} & 100 & 100 & 98.28 & \begin{tabular}[c]{@{}l@{}}Flower Framework 5 rounds with 10 clients.\end{tabular} \\ \hline
        \begin{tabular}[c]{@{}l@{}}Federated Learning (18 Features)\end{tabular} & 100 & 100 & 100 & \begin{tabular}[c]{@{}l@{}}Flower Framework 5 rounds with   10 clients.\end{tabular} \\ \hline
        Neto et al.~\cite{neto2024ciciov2024} & 95 & 95 & 95 & Random Forest and DNN. \\ \hline
        Yagiz et al.~\cite{yagiz2024transforming} & 100 & 100 & 100 & \begin{tabular}[c]{@{}l@{}} Developed a KD-XVAE model. Only attack data was used, \\while benign data was excluded, possibly leading to inaccurate evaluation results.
        \end{tabular} \\ \hline
        Mahdi et al.~\cite{mahdi2024advanced} & N/A & N/A & 99.77 & \begin{tabular}[c]{@{}l@{}}Focused on spoofing attacks only. \\ Used LSTM for classification and XGB for feature   selection.\end{tabular} \\ \hline
        Aswal et al.~\cite{aswal2024advancing} & 99.99 & 99.99 & 99.99 & \begin{tabular}[c]{@{}l@{}}Used CNN on   decimal data. \\ Did not use the full dataset, making results unverifiable.\end{tabular} \\ \hline
        Gul et al.~\cite{gul2024improving} & N/A & N/A & 99.64 & Used RF for the decimal dataset. \\ \hline
        %Naga et al.~\cite{naga2024boosting} & N/A & N/A & 99.49 & Flat classification (Gradient Boosting). \\ \hline
        
    \end{tabular}
\end{table*}

\subsection{Scalability and Overhead Analysis}

\begin{figure}[h]
       \centering
        \includegraphics[width = 0.85\linewidth]{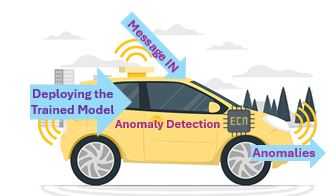}
        \caption{Overhead of the proposed approach}
        \label{fig:overhead}
        \vspace{-20pt}
\end{figure}

{\color{black}
To evaluate the scalability of the proposed approach, we analysed the overhead introduced at different layers of deployment, as illustrated in Figure~\ref{fig:overhead}. Training is performed in the cloud, where resource availability makes its cost negligible. The main overheads—evaluated using real-world parameters from Table~\ref{tab:networkconfiguration} and the configuration proposed by \cite{li2025impact}—relate to model deployment, memory usage, message processing, and additional network traffic for forwarding suspicious messages. Our experimental results show minimal impact on vehicle performance: memory usage is only 13 KB, and response time increases by just 0.13\%. With a packet delivery interval of 1 second, hourly model updates also generate negligible traffic. These results confirm the lightweight and scalable nature of the proposed model, which we aim to evaluate further through network simulations in future work.
}

\begin{table}[]
    \caption{Network configuration}
    \label{tab:networkconfiguration}
    \centering
\begin{tabular}{|l|l|}
\hline
\textbf{Parameter} & \textbf{Value} \\ \hline
Density& 180 veh/km\\
Minimum Distance& 2 m\\
Packet Delivery Interval & 1 s\\
Response Time & 1.5 s\\
Model Size & 13 KB\\
Model Update & 3600 s\\
Testing Time (Instance) & 0.002 s\\
\hline
\end{tabular}
\end{table}

\subsection{Computational Complexity}

 {\color{black}    
        In our proposed framework, classifiers are based on RF so that the complexity of our approach can be derived from the computational complexity of the training and prediction phases of RF.
            Specifically, the training complexity of RF depends on the complexity of tree construction, which is typically $\mathcal{O}(M \times N \times d)$.
            Here, $M$ is the number of features, $N$ is the number of training samples, and $d$ is the depth of the tree.
            When building a tree, as we recursively split the dataset at each node based on feature values to create child nodes, the depth of the tree grows.
            In a balanced binary tree, the maximum depth is approximately $\log N$. 
            The process continues until a stopping criterion is met, such as reaching a maximum depth (e.g., $\log N$), a minimum number of samples per leaf, or no further information gain. 
            Since an RF builds multiple decision trees (e.g., $T$ trees), the overall training complexity becomes $\mathcal{O}(T \times M \times N \times \log N)$.

        To make a prediction, we traverse each of the $T$ trees in the forest. 
            The complexity of traversing a single tree is $\mathcal{O}(d)$, where $d$ is the depth of the tree. 
            For a balanced tree, $d \approx \log N$, where $N$ is the number of training samples.
            Additionally, each node evaluation involves accessing and comparing a feature, contributing to the complexity.
            Thus the complexity for traversing a single tree is $\mathcal{O}(M \times d)$.
            Recall that in our proposed framework, the models are trained an then be deployed at different layers. 
            As such, the inference complexity of RF is equivalent to the prediction complexity. 
            Since the proposed framework consists of three layers, each with a RF classifier, the prediction complexity of one data point is $\mathcal{O}\big((T_1 \times \log N_1 + T_2 \times \log N_2 + T_2 \times \log N_2 + T_3 \times \log N_3) \times M \big)$, where $T_i \text{ and } N_i \text{ with}~i = {1, 2, 3}$ are respectively the number of trees and training samples at each layer.}

\subsection{Potential Overfitting Issue}

\textcolor{black}{ While analyzing the results from the RF and federated learning setting (as presented in Table 2, which is the Table IV in the revised manuscript), we observed that model accuracy improves consistently with the increase in the number of selected features. For example, specifically, with 10 features, we achieve 97.01\% accuracy for the 2-class task, 96.91\% for 3-class, and 95.23\% for 6-class classification in federated learning setting. When the number of features is increased to 11, the performance further improves to 100\% accuracy for both 2-class and 3-class, and 98.28\% for 6-class classification. Notably, with 18 selected features (out of 152), the model achieves 100\% accuracy across all class settings (2-class, 3-class, and 6-class). While this strong performance demonstrates the model's effectiveness in capturing discriminative patterns in the data, it also raises a legitimate concern regarding possible overfitting-particularly when perfect accuracy is observed. Overfitting might occur if the model memorises patterns in the training data rather than generalizing to unseen data. However, in our case, the improvements appear consistent and gradual as the number of features increases, which suggests that the selected features are informative and relevant to the classification task. To ensure that this behavior is not merely due to data leakage or poor generalization, we performed 10-fold cross-validation across all clients in the federated setup. Moreover, the selected features were chosen carefully using a wrapper-based selection strategy, and the FL training was done across distributed clients with non-overlapping data. Nonetheless, we acknowledge that achieving 100\% accuracy in a real-world federated setting is unusual and deserves further investigation. As such, we plan to extend this work by evaluating on more heterogeneous client datasets and integrating techniques such as differential privacy, client-level regularization, and dropout to enforce better generalization and prevent potential overfitting. }

\section{Discussion and Insights}

% Summarize the key takeaways from the results:
% The impact of feature selection on accuracy.
% The superiority of your model compared to existing approaches.
% The benefits of the hierarchical structure in classifying IoV attacks.
% Address any limitations in the results and suggest potential directions for future research.

The proposed three-level hierarchical classification framework for multilayer IoV networks addresses the critical challenges of efficiency, scalability, and privacy in IoV systems. 
    The hierarchical design ensures that each classification level is strategically deployed: the first level at the vehicle for quick identification of benign or malicious instances, the second level at the roadside unit for categorizing attacks, and the third level at the edge network for detailed attack-specific classification. This division of responsibilities allows for localized decision-making, reducing the computational burden on individual components. A notable benefit of the proposed framework is its lightweight nature, achieved through the combination of Boruta feature selection and SHAP explanations. By reducing the feature set to the most relevant attributes, the model ensures efficient resource usage while maintaining high accuracy. For instance, using only 11 features (out of 152) achieves perfect classification for two-class and three-class problems, while 18 features are sufficient for six-class classification. This feature reduction is essential for IoV networks, where resource constraints and real-time processing requirements are critical. 
    The use of a Deep Neural Network (DNN) in the FL setting further enhances the framework by enabling privacy-preserving training across distributed nodes. This approach ensures data security and aligns with the privacy requirements of IoV networks, while the centralized cloud server facilitates model training and retraining when necessary. Lightweight retraining capabilities at the roadside unit and edge nodes provide additional flexibility for dynamic IoV environments.
The analysis demonstrates the superior of our proposed framework compared with existing approach, thanks to the effective feature selection component in our solution.  
% that the CIC-IoV2024 dataset is easier for the model to learn compared with other datasets in the literature, achieving 100\% accuracy for all classification levels using appropriate feature selection and hierarchical classification. 
This result significantly outperforms the 95\% accuracy reported in the baseline study~\cite{neto2024ciciov2024}. 
Furthermore, unlike existing works~\cite{yagiz2024transforming, mahdi2024advanced, aswal2024advancing, gul2024improving} that often rely on flat classification with all features, the proposed hierarchical approach aligns better with the decentralized nature and resource constraints of IoV networks. 
Note that while some works (e.g.,~\cite{yagiz2024transforming}) claim 100\% accuracy, their reliance on flat classification and requiring hundred features make them less practical for real-world IoV applications.

%==================================================
\section{Conclusions and Future Work}
\label{Conclusion}

This work proposed a scalable, three-level hierarchical classification model for intrusion detection in Internet of Vehicles (IoV) networks, utilizing the CIC-IoV2024 dataset. Our model was designed to address the unique challenges of IoV, including resource constraints and the need for fine-grained attack classification. By employing Boruta Feature Selection, we can significantly reduce the computational complexity of training and testing while maintaining high classification performance. 
The hierarchical structure provided multi-level granularity in attack detection, distinguishing between benign and attack traffic at Level 1, categorizing coarse-grained attack types at Level 2, and identifying specific attack subtypes at Level 3.

The model was rigorously evaluated using 10-fold stratified cross-validation with various machine learning algorithms. Results demonstrated that the hierarchical approach significantly outperformed traditional flat classification models, achieving higher accuracy, precision, recall, and F1-scores. The comparative analysis also underscored the scalability and adaptability of our method across diverse attack scenarios.

Despite its effectiveness, our study highlights areas for future improvement, including extending the model to incorporate real-time detection capabilities, testing on additional IoV datasets, and exploring deep learning approaches for further optimization. From our experiment, we observed that the FL-based approach yielded good results. Therefore, our future work is to integrate the proposed hierarchical classification framework into a federated learning setting. This advancement would combine the benefits of hierarchical granularity with privacy and security inherent in federated learning. Such an approach could enable scalable, secure, and efficient classification systems for IoV networks, further enhancing their adaptability to evolving cyber threats.

\textcolor{black}{Beyond the evaluation conducted in this study, future research will focus on validating the proposed approach across multiple datasets. 
        While CIC-IoV2024 provides a comprehensive foundation for IoV security analysis, we plan to extend our experiments to datasets such as CIC-IDS-2018 to assess the method’s generalizability in different network environments. 
        Additionally, as existing IoV security datasets lack privacy attacks, we aim to explore this domain in our future work to fill this important gap in the literature.  
        Moreover, we also plan to implement it in a real-time environment to assess its performance under live network conditions, considering factors such as latency, adaptability, and scalability.}

\textcolor{black}{
While this study focuses on improving intrusion detection through hierarchical classification, future research could explore the integration of networking aspects, such as adaptive traffic routing and network-aware anomaly detection, to enhance the overall security and efficiency of IoV systems. Investigating how communication delays and bandwidth constraints impact the deployment of hierarchical IDS in real-world IoV environments would be a valuable extension of this work.}

%==================================================
% \section*{Declarations}

% \subsection*{Conflict of interest}
% The authors have no conflicts of interest to declare that they are relevant to the content of this article.

% \section*{Acknowledgments}

% \section*{Author statements}
% XYZ: Conceptualization; Data curation; Implementation, Writing-original draft; and Writing, Visualization; Formal analysis.
% XYZ: Funding acquisition; Investigation; Methodology; Project administration; Resources; Software; Supervision; Validation; Writing-original draft; and Writing - review \& editing.
% XYZ: Investigation, Visualization; Project Administration.
% XYZ: Investigation, Visualization.
% XYZ: Data curation; Implementation; Investigation, Visualization; Validation, Review \& editing.

\bibliographystyle{IEEEtran}
\bibliography{ids-iov}

% {\appendix[Proof of the Zonklar Equations]
% Use $\backslash${\tt{appendix}} if you have a single appendix:
% Do not use $\backslash${\tt{section}} anymore after $\backslash${\tt{appendix}}, only $\backslash${\tt{section*}}.
% If you have multiple appendixes use $\backslash${\tt{appendices}} then use $\backslash${\tt{section}} to start each appendix.
% You must declare a $\backslash${\tt{section}} before using any $\backslash${\tt{subsection}} or using $\backslash${\tt{label}} ($\backslash${\tt{appendices}} by itself
%  starts a section numbered zero.)}


\newpage

 

% \end{IEEEbiography}



\vfill

\end{document}